\documentclass[11pt,twoside,a4paper]{article}
\usepackage{fancyhdr}
\usepackage{url}
\newcommand{\cancel}[1]{}

\usepackage{cite}
\usepackage{amsmath,amssymb,amsfonts}
\usepackage{algorithmic}
\usepackage{graphicx}
\usepackage{textcomp}
\usepackage{makecell}
\usepackage{hyperref}
\usepackage{url}

\usepackage{multirow}
\usepackage{caption,setspace}
 
\newtheorem{researchproblem}{\bf Research Problem}

\def\BibTeX{{\rm B\kern-.05em{\sc i\kern-.025em b}\kern-.08em
    T\kern-.1667em\lower.7ex\hbox{E}\kern-.125emX}}

\begin{document}

\title{SoK of Used Cryptography in Blockchain}

\author{Mayank Raikwar\thanks{Department of Information Security and Communication Technologies, Norwegian University of Science and Technology - NTNU} \and Danilo Gligoroski$^*$ \and Katina Kralevska$^*$}

\maketitle

\begin{abstract}
The underlying fundaments of blockchain are cryptography and cryptographic concepts that provide reliable and secure decentralized solutions. Although many recent papers study the use-cases of blockchain in different industrial areas, such as finance, health care, legal relations, IoT, information security, and consensus building systems, only few studies scrutinize the cryptographic concepts used in blockchain. To the best of our knowledge, there is no Systematization of Knowledge (SoK) that gives a complete picture of the existing cryptographic concepts which have been deployed or have the potential to be deployed in  blockchain. In this paper, we thoroughly review and systematize all cryptographic concepts which are already used in blockchain. Additionally, we give a list of cryptographic concepts which have not yet been applied but have big potentials to improve the current blockchain solutions. We also include possible instantiations of these cryptographic concepts in the blockchain domain. Last but not least, we explicitly postulate 21 challenging problems that cryptographers interested in blockchain can work on. 
\end{abstract}

\newpage
\tableofcontents

\newpage
\section{Introduction}
Blockchain, a distributed ledger managed by a peer-to-peer network collectively adhering to some consensus protocol, is arguably considered as a new and disruptive technology. Both academia and industry are profoundly affected by new solutions to some old problems which are based on this new technology. The success of the blockchain concept is ultimately connected with the financial success of Bitcoin~\cite{Nakamoto_bitcoin} that was developed just one decade ago, and the subsequent avalanche of more than 2140 other crypto-currencies that all together built a financial market worth around \$285 billion (as of 16 June 2019) \cite{coinmarketcap}. 

We can trace the origins of the ideas to use cryptography for secure and private transactions for paying access to databases, paying for services such as online games, transferring money over the Internet, Internet shopping and other commercial activities back in 1990's with David Chaum's eCash system \cite{e-cash}. One of the negative aspects of eCash was that it was a centralized system, controlled by a trusted third party. Another hurdle for a broader acceptance of eCash was the fact that it was covered by a long list of patented algorithms -- something that is considered as a big obstacle to acceptance among the crypto community.

In parallel, in 1990's we saw the development of several cryptographic ideas not directly connected but somehow still related to the ideas of using cryptography in financial transactions. We mention some of them such as the proposal on how to combat junk email \cite{dwork1992pricing} by Dwork and Naor that was published in 1992, and which used computationally expensive functions. Then in 1996, there was a proposal for time-lock cryptographic puzzles \cite{Rivest:1996:TPT:888615} by Rivest, Shamir, and Wagner by using RSA based CPU expensive computations. At the end of 90's and early 2000's several patent free cryptographic concepts were proposed, implemented and released as open source projects by an online movement and a community of cryptographers and programmers known as  ''Cypherpunks'' \cite{hughes1993cypherpunk}. Those cryptographic concepts and implementations include Adam Back's ''hashcash'' proposal for a currency based on the hardness of finding partial hash collisions \cite{back2003hashcash}, Wei Dai's ''b-money'' \cite{dai1998b} and Nick Szabo's\footnote{Nick Szabo was also part of the eCash development team in late 90's.} ''Bitgold'' proposal \cite{Szabo2005}. These concepts have been the basis of the Satoshi Nakamoto's decentralized cryptocurrency, nowadays known as Bitcoin ~\cite{Nakamoto_bitcoin,satoshi_comment}. As a recognition of their pioneering activities in the decentralized cryptocurrencies, Ethereum~\cite{Ethereum} -- the second most popular cryptocurrency -- named the three of its denominations as "Wei", "Szabo" and "Finney"~\cite{EthereumDenominations}.\footnote{Hal Finney was a cypherpunk and the receiver of the first Bitcoin transaction of 10 Bitcoins from the anonymous Satoshi Nakamoto~\cite{HalFinney}.}

The underlying core technology in Bitcoin is blockchain. Blockchain is a distributed ledger maintaining a continuously growing list of data records that are confirmed by all of the participating nodes. The data is recorded in this public ledger in a form of blocks of valid transactions, and this public ledger is shared and available to all nodes.

\par Blockchain is envisioned as a promising and powerful technology but it still encounters many research challenges. Some of the main challenges are constant improvement of its security and privacy, key management, scalability, analysis of new attacks, smart contract management, and incremental introduction of new cryptographic features in existing blockchains. These challenges arise due to the network structure and the underlying consensus mechanisms and cryptographic schemes used within the blockchains. To overcome these challenges and to find enhanced solutions, many of the cryptographic concepts such as signature schemes, zero-knowledge proofs, and commitment protocols are scrutinized and applied. As cryptography is a vast research field, there is always a scope to find new cryptographic schemes in order to improve the solutions in blockchain.\\
The majority of the ongoing research in Blockchain focuses on finding and identifying improvements to the current processes and routines, mostly in industries that rely on intermediaries, including banking, finance, real estate, insurance, legal system  procedures, and healthcare. The study on business innovation through blockchain~\cite{Usecases} presents some blockchain enabled business applications and their instantiations. These blockchain enabled applications still need a proper way for selecting the cryptographic technique employed in their respective solution in order to meet the business requirements. Not only these blockchain applications but also the research community will benefit from an overview in a form of systematization of the current state of knowledge of all available cryptographic concepts which have been applied or can be applied in existing and future blockchain solutions. 
To the best of our knowledge, this is the first systematization of knowledge that gives a complete picture of the existing cryptographic concepts related to blockchain. We have tried to depict most of the cryptographic concepts in the blockchain domain. Although there are various works about specific cryptographic concepts used in blockchain, there are only few works which merge all these atomic works and present them in a single paper. Most of the review and survey works such as~\cite{S&P,ConsensusSurvey} discuss security, privacy, consensus or other challenges in blockchain. A recent work of Wang et al.~\cite{CryptoPrimitives} gives a comprehensive analysis of cryptographic primitives in blockchain. Their analysis presents the functionality and the usage of these primitives in blockchain. However, the analysis is based only on existing cryptocurrencies and it lacks many of the cryptographic protocols which are used in blockchain.


\subsection{Our Contribution}
In this study, we classify cryptographic concepts based on their use in blockchain. We have divided them into two categories: 1. Concepts which are well used in blockchain, and 2. Concepts which are promising but not yet implemented in blockchain.  This categorization does not have a clear boundary. We classify some cryptographic concepts as promising ones, and that requires further research and scrutiny in order to be deployed in blockchain. As a result, the following points are the main contributions of our Systematization of Knowledge (SoK) paper:
\begin{itemize}
    \item We provide a description of cryptographic concepts which have been applied in the blockchain field. We also include instantiation of these concepts in blockchain. 
    \item We provide a list of cryptographic concepts which are rarely used or have not been used in blockchain but they have the potential to be applied in this field. These concepts open many possible research directions and they can be examined in different blockchain applications.
    \item We identified 21 research challenges that we formulate as \textit{Research Problem}. Some of them are rephrased research challenges already published in the literature and some of them are newly formulated research problems.
\end{itemize}

In this study, we do not claim that we have exhausted all of the cryptographic concepts which are employed in blockchain, but we have tried to cover the concepts which we felt are propitious for the blockchain domain. We also describe each cryptographic concept along with its associated properties and its instantiation in the blockchain field. Additionally, in order to give one unified presentation about blockchain, we give a brief explanation about:

\begin{itemize}
    \item Enabling concepts of blockchain such as hash function, consensus protocol, network architecture.
    \item Layered architecture of blockchain and emphasis on some of the major challenges associated with blockchain.
\end{itemize}



\section{Research Methodology} \label{Methodology}
To perform a systematization of knowledge of the existing cryptographic concepts related to blockchain, we established and followed a methodology that we explain in this Section. Since the invention of Bitcoin, there has been a growing interest in blockchain from both academia and industry. The number of publications in the blockchain field has been rapidly increasing in recent years. Not all of these publications are research works; some of these works discuss different use-cases of blockchain. Therefore, to review these many papers in the blockchain field, we pursued a research methodology which defines the inclusion criteria, a search strategy to search for respective publications and a data collection mechanism to accumulate the relevant publications. The collected data is later processed based on inclusion and exclusion criteria. The publications which meet the inclusion criteria go through one final step of quality assessment. Once a publication passes the quality assessment, it is included in our systematization.

We use keyword search to make the first selection of potentially relevant scientific publications. For the keyword search, we typed keywords such as $\ll \textit{cryptographic concept name} \gg \ll\textit{in blockchain}\gg \mathrm{or} \ll\textit{use of}\gg \ll\textit{cryptographic concept name}\gg \ll\textit{in blockchain}\gg$. We use Google Scholar as our primary source to search for the relevant literature, but as Google Scholar does not exhaust all of the available literature, we also searched in databases such as: 1) IACR eprint archive, 2) IEEE Xplore, 3) ACM Digital Library, 4) ScienceDirect, and 5) Springer Link. 

\par 
The inclusion criteria for this study is based on the following questions:
\begin{itemize}
    \item Is the elaborated cryptographic concept useful in blockchain? The usefulness of the cryptographic concept is measured as whether we achieve some essential properties in blockchain by using the concept or whether the cryptographic concept can be beneficial for some use-case compared to an already implemented concept.
    \item Which properties can be achieved by using the cryptographic concept in blockchain?
    \item Is there any instantiation of the cryptographic concept in a blockchain study or application? If not, is there any potential?
\end{itemize}

The criteria for excluding a paper is:
\begin{itemize}
    \item Informal literature discussing some cryptographic concepts in blockchain.
    \item Literature which claims on using a cryptographic concept but it does not give any guarantees about the feasibility and prospects of a potential implementation.
\end{itemize}

The quality of the papers that meet the inclusion criteria is assessed. For quality assessment, we apply the following questions:
\begin{itemize}
   
    \item Is the cryptographic concept implemented in blockchain? If not, is it possible to implement it and will it be more efficient than the existing solution?
    \item Is there any security analysis or does the implemented concept rely on another underlying platform?
    \item Are the fundamental concept and its related properties adequately described?
\end{itemize}

 \section{Supporting and enabling concepts of Blockchain} \label{Blockchain-basic}
\begin{figure}[htbp]
     \centering
     \includegraphics[width = 0.75\textwidth]{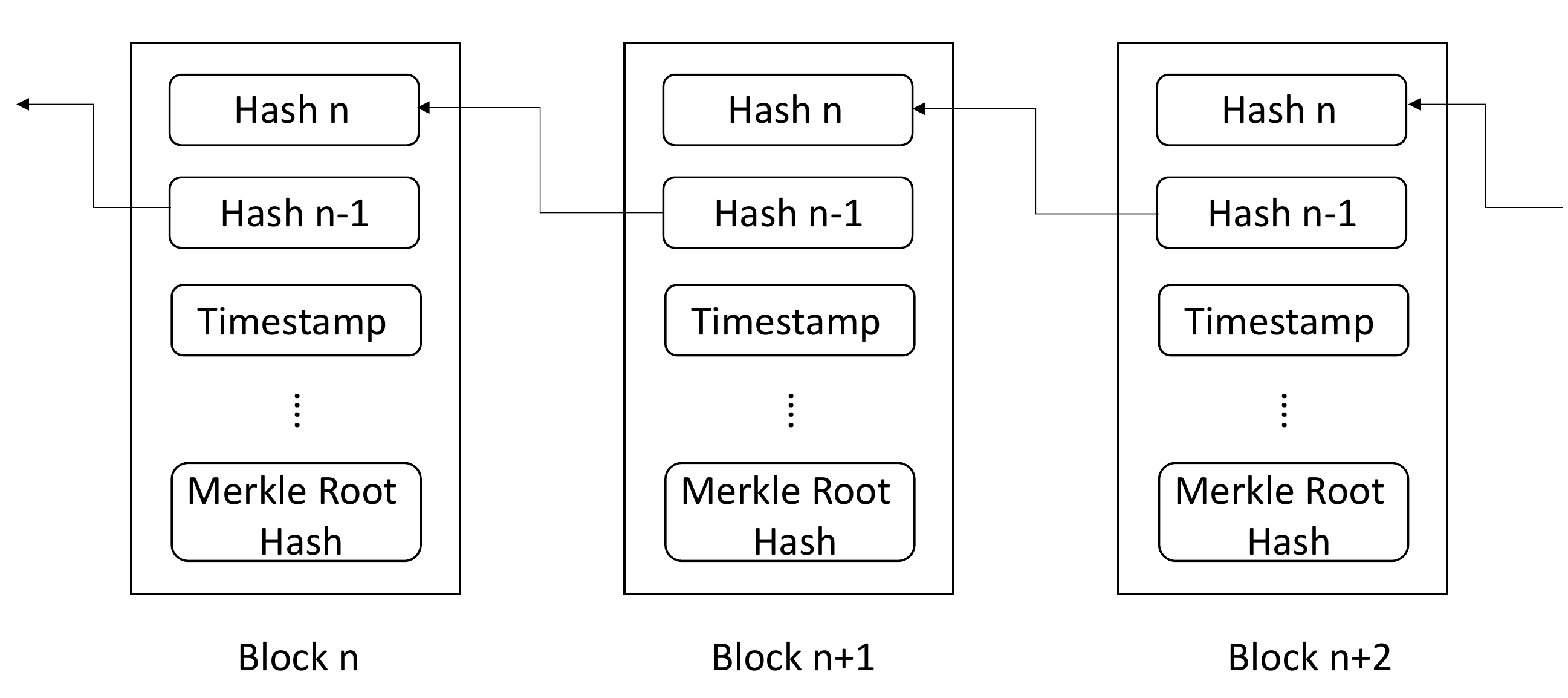}
     \caption{Basic blockchain structure.}
     \label{Blockchain}
 \end{figure}

\begin{figure*}[htbp]
    \centering
    \includegraphics[width = \textwidth]{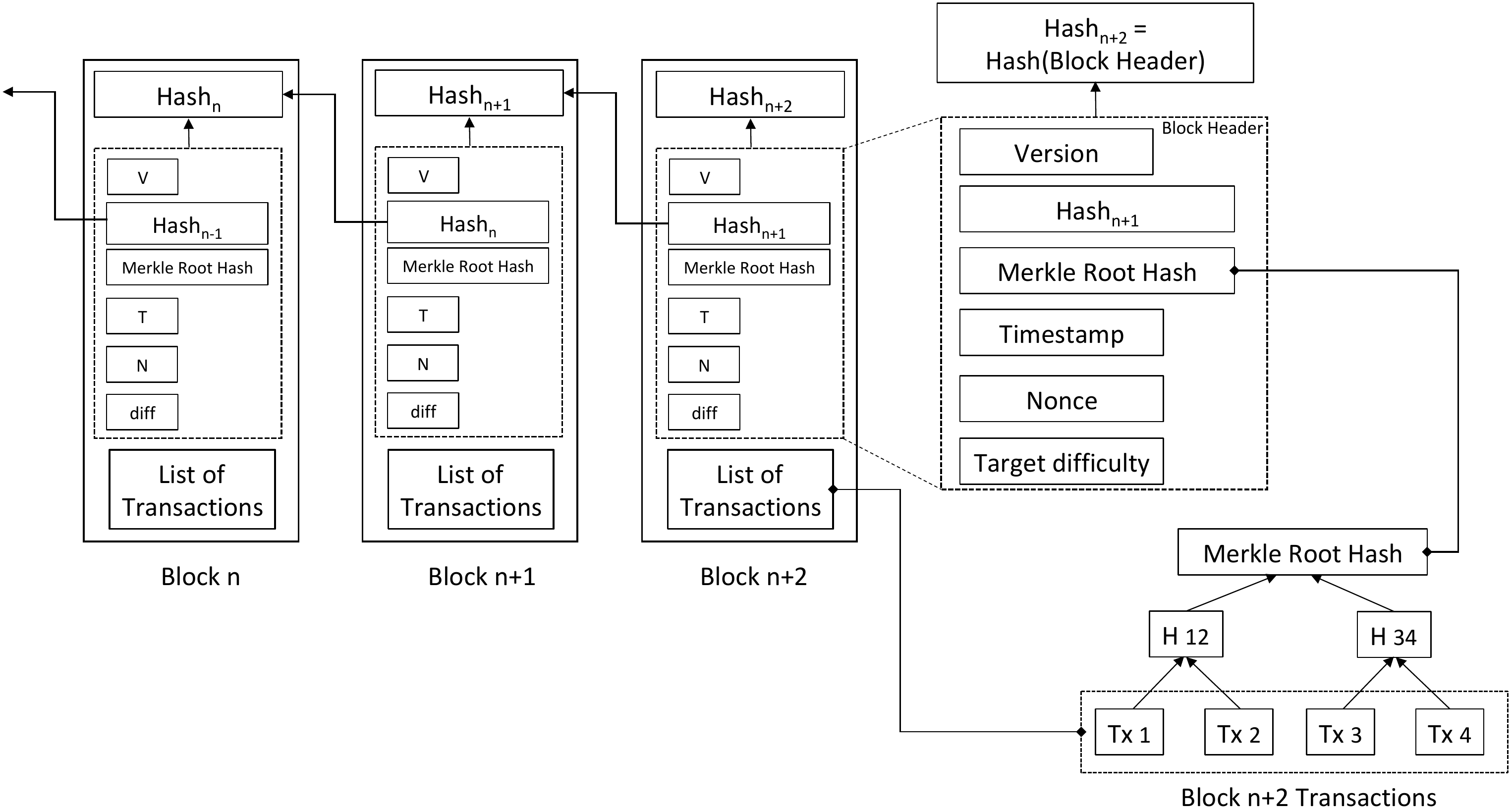}
    \caption{Blockchain data structure with block format.}
    \label{Blockchain-Data-Structure}
\end{figure*}

 As previously mentioned, blockchain is a way to encapsulate transactions in the form of blocks where blocks are linked  through the cryptographic hash, hence forming a chain of blocks. Figure~\ref{Blockchain} shows the basic blockchain structure. Each block in the blockchain contains a block header and a representation of the transaction. For instance, in Figure~\ref{Blockchain}, each block consists of its hash, the hash of the previous block, a timestamp and some other block fields (e.g., version, nonce). This depends from the block design. Merkle root hash represents the set of transactions in the Merkle tree, and this representation of transactions varies according to the design of the blockchain implementation. Figure~\ref{Blockchain-Data-Structure} depicts the Bitcoin blockchain data structure showing in details the block format.
 
Blockchain relies on different constituents which serve different purposes. In this Section, we give an overview of the main underlying concepts used to build a blockchain. A detailed technical explanation of all these concepts is out of the scope of this paper, but we have tried to cover the essentials of their functionality.

\subsection{Cryptographic Hash Function} \label{Hash Function}

A hash function $H$ is a function which takes an input of an arbitrary size and maps it to a fixed size output. Cryptographic hash functions have some additional properties such as: \textbf{a}) \textit{collision resistance} - it is hard to find two inputs $a$ and $b$ such that $H(a) = H(b)$; \textbf{b}) \textit{preimage resistance} - for a given output $y$ it is hard to find an input $a$ such that $H(a) = y$; and \textbf{c}) \textit{second preimage resistance} - for a given input $a$ and output $y = H(a)$ it is hard to find a second input $b$ such that $H(b) = y$. Readers interested in an extensive cover of the field of cryptographic hash functions are referred to~\cite{preneel1998state}.
 
 Cryptographic hash functions in blockchain are used for various purposes such as: 
 \begin{enumerate}
     \item solving cryptographic puzzles (the Proof of Work (PoW) in Bitcoin~\cite{Nakamoto_bitcoin});
     \item address generation (for public and private keys);
     \item shortening the size of the public addresses;
     \item message digests in signatures.
 \end{enumerate}
  
The most popular cryptographic hash functions used in blockchains are SHA-2~\cite{gallagher1995secure} (especially the variant SHA256 - a variant that produces outputs of 256 bits), and some of the well analyzed hash functions from the NIST SHA-3 competition and standardization that went to the later stages of that process (final 5 proposals or some of the 14 proposals from the second phase~\cite{regenscheid2009status}). Some of the existing blockchain designs such as IOTA constructed their own ''homebrewed'' cryptographic hash function called Curl-P, that was received very critically and negatively by the crypto community~\cite{heilman2019cryptanalysis,heilman2017iota}. 

A typical way how cryptographic hash functions are used in blockchain designs is in a form of a mode of operation, i.e., a combination of several invocations of a same or different hash functions. For example, in 
Bitcoin~\cite{Nakamoto_bitcoin}, SHA256 is used twice and that construction is called $\mathrm{SHA256d}$, i.e., \begin{equation}\small \mathrm{SHA256d}(message) = \mathrm{SHA256}( \mathrm{SHA256}(message) ). \end{equation}

$Mining$ is a process of creating a new block of transactions through solving a cryptographic puzzle, and the participant who solves the puzzle first is called a $miner \ of \ the \ block$. 
If we look at the Bitcoin PoW puzzle, we can see that a miner has to find a $Nonce$ (similar to Hashcash protocol~\cite{back2003hashcash} that we discuss in the next subsection) to create the next block in the blockchain. The puzzle looks like this:
    \begin{equation}\label{PoW-Puzzle}\small
        \mathrm{SHA256d}(Ver || HashPrevBlock ||\ldots|| Nonce) \leq T
    \end{equation}
where $T$ is 256-bit target value.
    
    \begin{table}[h]
        \centering
        \resizebox{0.80\textwidth}{!}{
        \begin{tabular}{c|c}
            Target Value $T$ & Fraction of SHA256d outputs $\leq T$\\
            \hline
          $0\texttt{x7} \underbrace{\mathtt{t_{1}t_{2}t_{3}t_{4} \ldots t_{62}t_{63}}}_{63\,times}$  & \Large${\frac{1}{2}}$ \\
          $0\texttt{x0} \underbrace{\mathtt{t_{1}t_{2}t_{3}t_{4} \ldots t_{62}t_{63}}}_{63\,times}$  & \Large$\frac{1}{16}$ \\
          $0\texttt{x}\underbrace{\mathtt{00\ldots00}}_{16\,times} \underbrace{\mathtt{t_{1}t_{2}t_{3}t_{4} \ldots t_{47}t_{48}}}_{48\,times}$ & \Large$\frac{1}{2^{64}}$\\
        \end{tabular}}
        \caption{Fraction of SHA256d outputs with respective target value.}
        \label{minig}
    \end{table}

Looking into the fraction of $\mathrm{SHA256d}$ outputs that are less than the target value $T$ for different values of $T$ in Table~\ref{minig} helps us to understand why mining is hard in PoW. Namely, the probability of finding a nonce that will cause the whole block to have a hash that is less than the target value is 
\begin{equation}
    Pr[\mathrm{SHA256d}(Block) \leq T] \approx \frac{T}{2^{256}}.
\end{equation}

We next discuss the research and innovative activities in the area of cryptographic hash functions that were either remotely or directly connected and inspired by the trends in blockchain.

Several years after the launch of the Bitcoin and its source code being published as an open source on Github, blockchain designers started to clone and fork its basic code, and started to introduce different variants and innovations. One of the earliest forks from 2011 that is still popular nowadays is Litecoin~\cite{Litecoin}. The basic idea by the Litecoin design was to use a different hash function for its proof of work puzzles. The motivation came from the fact that even in 2011 there were trends to build specialized application-specific integrated circuit (ASIC) hardware implementations of $\mathrm{SHA256d}$ that will mine the blocks several orders of magnitude faster than ordinary CPUs and GPUs. Instead of $\mathrm{SHA256d}$, Litecoin uses $Scrypt$ \cite{percival2009stronger} - a memory-intensive compilation of use of the HMAC \cite{krawczyk1997hmac} construction instantiated with SHA256 and use of the stream cipher Salsa20/8 \cite{bernstein2008salsa20}. The idea was that the use of $Scrypt$ will be impractical to implement it in ASIC, thus, giving chances of individual owners of regular computers and GPUs to become a significant mining community. While with no doubts we can say that Litecoin is a very successful alternative cryptocurrency, we can for sure claim that its initial goal to be ASIC resistant blockchain design was not successful. Nowadays, you can find commercial products for Litecoin hardware mining\footnote{One such a product that can compute 580 billion Scrypt hashes per second, is offered by the company Bitmain and is called "Antminer L3++". As of the time of writing this article, this product was advertised at \break  \url{https://shop.bitmain.com/} for a price of \$213.00 and for a 10 days delivery (2 June 2019).}. 

Actually, we can say that the 10 years of history of blockchain, in general, and cryptocurrencies, in particular, is a history of failed attempts to construct a sustainable blockchain that will prevent the appearance of profitable ASIC miners that can mine the blocks with hash computing rates that are several orders of magnitude higher than the ordinary users of CPUs and GPUs. In that short history, we can mention Ethash used in Ethereum~\cite{Ethereum} for which there are now commercially available ASIC miners by at least two companies. In 2013, QuarkCoin~\cite{Buterin2013} introduced the idea of using a chain of six hash functions (five SHA-3 finalists BLAKE, Gr\o{}stl, JH, Keccak and Skein \cite{turan2011status}) and the second round hash function Blue Midnight Wish~\cite{gligoroski2009cryptographic}. One of the motivations behind the QuarkCoin PoW function was to be more ASIC resistant than $\mathrm{SHA256d}$. The cascading idea of QuarkCoin was later extended to a cascade of eleven hash functions in Darkcoin (later renamed DASH~\cite{duffield2018dash}). Needless to say, nowadays there are commercially available ASIC miners for X11 as well. 

The frictions between ASIC miners and the cryptocurrency community seem to remain to the present days, and are somewhat evolving and inspiring novel proposals in blockchain protocols. The latest is the Programmatic Proof-of-Work (ProgPoW) initiative for Ethereum blockchain ecosystem that aims to make ASIC mining less efficient and to give some advantages to graphics processing units (GPU) mining~\cite{ProgPoW2019}.

\subsection{Consensus Mechanisms}\label{ConsensusMechanisms}
Consensus is the key component of blockchain to synchronize or update the ledger by reaching an agreement among the participants. In order to maintain the ledger in a decentralized way, many consensus mechanisms have been proposed. The first introduction of the use of a consensus mechanism in blockchain is implicitly given by Bitcoin. Bitcoin uses Proof of Work (PoW) mechanism as consensus where the idea came from \textit{Hashcash Protocol}~\cite{back2003hashcash}. The objective of Hashcash was to prevent spam in public databases. The \textit{Hashcash Protocol} is as follows. Suppose an email client wants to send an email to an email server. In the beginning, the client and the server both agree on a cryptographic hash function $\mathit{H}$ which maps an input string to an $n$ length output string. Then, the email server sends a challenge string $\mathit{c}$ to the client. Now the client has to find a string $\mathit{x}$ such that $\mathit{H(c||x)}$ starts with $\mathit{k}$ zeros. Since $\mathit{H}$ has pseudorandom outputs, the probability of success in a single trial is 
    { \large
    \begin{center}
        $\frac{2^{n-k}}{2^n} = \frac{1}{2^k}.$
    \end{center}}
    

Here $\mathit{x}$ corresponding to $\mathit{c}$ is considered as PoW and the process of finding that $\mathit{x}$ is called mining. PoW is difficult to generate but easy to verify.

Many literature studies on consensus mechanisms, for instance, the survey by Wang et al. ~\cite{ConsensusSurvey} and "SoK: Consensus in the age of blockchains"~\cite{SoK}, have been carried out in the past few years. Since consensus mechanisms have already been thoroughly studied in the literature, in this paper, we present the basic idea about how consensus mechanisms work and their classification.

In a consensus protocol, depending on the network architecture and blockchain type, some or all of the participants take part and maintain the ledger by adding a block consisting of transactions to their ledger. However, the creation of a new block to be added to the ledger is performed by a participant who is known as a leader of the consensus protocol in that particular execution. This leader is elected by different mechanisms of leader election process, and some of these mechanisms are given in Table~\ref{Leader Election}.  
        \begin{table}[h]
            \centering
            \resizebox{0.7\textwidth}{!}{
            \begin{tabular}{|l|c|}
            \hline
            
              \textbf{Leader Election Criteria}  & \textbf{Reference Protocols} \\
              \hline
              PoW Puzzle Competition  & \makecell{Bitcoin-NG~\cite{Bitcoin-NG}, Casper~\cite{Casper},\\ Proof of Stake velocity~\cite{POSV}} \\
              \hline
              Verifiable Random Function & \makecell{Tendermint~\cite{tendermint}, Algorand~\cite{Algorand},\\ Secure Proof of Stake~\cite{POS}} \\
              \hline
              Trusted Random Function & \makecell{Proof of luck~\cite{POL},\\ Proof of elapsed time~\cite{POET}}\\
              \hline
              Modified Preimage Search & Snow White~\cite{Snow-White}\\
              \hline
              \makecell{Sub-network of\\ Masternodes / Validator nodes} & Darkcoin  and DASH~\cite{duffield2014transaction}, Libra~\cite{Libra} \\
              \hline
            \end{tabular}}
            \caption{Leader election in consensus protocols.}
            \label{Leader Election}
        \end{table}

 After the leader is elected and the new block is created in order to achieve consensus or agreement on this block, two types of voting mechanisms are followed: \textit{explicit} and \textit{implicit}. In explicit voting, multiple rounds of voting occur and then based on the votes, consensus is reached. However, in implicit voting, the new block created by the leader is accepted by others who implicitly vote for the new block and add it to their ledgers. A leader election through PoW puzzle competition (e.g., PoW puzzle~\ref{PoW-Puzzle} in Bitcoin) followed by an implicit voting to reach an agreement is also called "Nakamoto Consensus".  
\begin{figure}[ht]
    \centering
    \includegraphics[width = 0.75\textwidth]{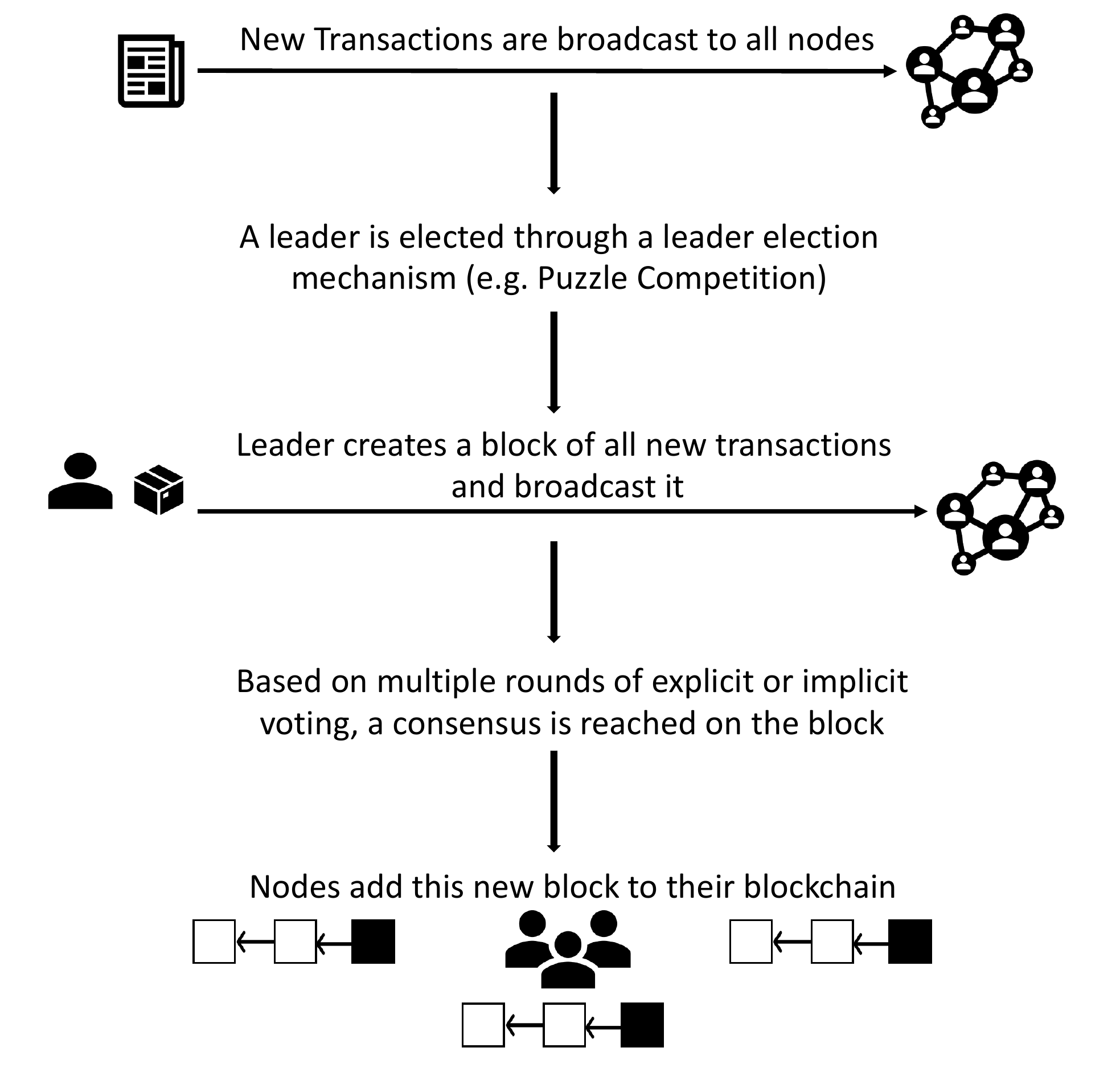}
    \caption{Blockchain consensus scenario.}
    \label{Consensus}
\end{figure}

Consensus mechanisms also determine the performance of the blockchain network in terms of consensus finality, throughput, scalability, and robustness against various attacks. In some manner, consensus orchestrates the state of the programs executed in the blockchain network nodes by providing a runtime environment to collectively verify the same program and hence reach to a finality. There is no exact classification of consensus mechanisms, but in general they can be classified as consensus protocols with proof of concept and consensus protocols with byzantine fault-tolerant replication. These consensus protocols can be chosen based on the blockchain network and type. Most of the proof of concept consensus protocols are used in permissionless blockchains. There are many proof of concept schemes which have been proposed and implemented, e.g., Proof of Work (PoW)~\cite{PoW}, Proof of Stake (PoS)~\cite{Proof-of-stake}, Equihash~\cite{Equihash}, having Masternodes in Dash~\cite{duffield2014transaction},  etc. As described in Section~\ref{Hash Function}, in PoW puzzle based consensus protocols, miners try to solve the cryptographic puzzle by mining and these miners are also responsible for verification of the transactions, and an incentive ($reward$) is given to the first miner who solves the puzzle. 

In case of a permissionless network, as there is no authentication and no proper synchronization, the underlying consensus algorithm should be able to handle the synchronization problem, scale well and mitigate different attacks in order to maintain canonical blockchain state in P2P network. To solve this synchronization issue, most of the blockchains use "Longest chain rule" to have a consistent canonical state of blockchain in this P2P blockchain network. On the contrary, in the permissioned blockchain, as there are restrictions and privileges associated with the peers, there is a strict control on the synchronization among the peers. Byzantine fault-tolerant protocols are usually adopted in permissioned blockchains to provide consensus properties such as validity, agreement, and termination. Practical Byzantine Fault Tolerant (PBFT)~\cite{PBFT}, Proof of Elapsed Time~\cite{POET}, Ripple consensus~\cite{Ripple-consensus} are some of the consensus protocols used in permissioned blockchains. Recently, Facebook launched his own global cryptocurrency `Libra'~\cite{Libra} which works as a permissioned blockchain and provides users to do transactions with nearly zero fee. 
Libra blockchain comes with a new programming language `Move' and new consensus protocol `LibraBFT'.

\subsubsection*{Mining, Pool Mining and Incentive Mechanisms} In Proof of Work based blockchains, the addition of new transactions in the blockchain is performed by the mining process. In the Bitcoin mining process, a puzzle is solved by computing many hashes repeatedly (Equation~\ref{PoW-Puzzle}) by putting different values for the nonce to satisfy the condition. When a miner successfully solves the puzzle first among all of the miners, it gets a monetary incentive for solving the puzzle. Because of this incentive process, all consensus nodes or miners follow the rules of the blockchain state transition during the puzzle competition. Mining is a resource-intensive process where the main resources are computational power and memory. Mining can be performed either by a solo miner or by a group of miners, called a mining pool, who collectively try to solve the puzzle. Mining pools may operate on different mining techniques and incentive mechanisms. These incentive mechanisms can vary based on the used mining technique or the decision of the pool operator. Reference~\cite{ConsensusSurvey} gives a brief idea about the mining strategy management in blockchain networks, while reference~\cite{Mining-games} provides a strategic study of mining through stochastic games. Different incentive mechanisms are proposed and tested in blockchains. Reference~\cite{Bitcoin-reward-analysis} analyzes Bitcoin pooled mining reward systems, and a reward system based on information propagation in blockchain network is presented in~\cite{RedBallons}.

\subsection{Network Infrastructure}

Blockchain is maintained by a peer-to-peer (P2P) network. P2P network is an overlay network which is built on the top of the Internet. This P2P blockchain network can be modeled as structured, unstructured or hybrid based on several parameters such as the consensus mechanism and the type of blockchain. Regardless of the representation of the network, a blockchain network should quickly disseminate the newly generated block so that the global view of the blockchain remains consistent. Consequently, a synchronization protocol is needed, but a routing protocol might or might not be needed. A traditional P2P network uses a routing protocol to route the information through multihop; however, in many blockchains (e.g., Bitcoin), routing is not required because a peer can get information through at most one hop, so no routing table is maintained. 

Almost all cryptocurrencies and blockchains such as Bitcoin~\cite{Nakamoto_bitcoin}, Ethereum~\cite{Ethereum}, Litecoin~\cite{Litecoin} use unstructured P2P network where the idea is to have equal privileges for all of the nodes and to create an egalitarian network. A P2P network can follow flat or hierarchical organization for building a random graph among the peers. This graph is not fully connected, but in order to receive all of the communication and to maintain the ledger, each peer maintains a list of peer addresses. Thus, if any peer propagates a message in the network, eventually all peers receive it through their available connections. In an unstructured network, techniques like flooding and random walk are used to make new connections with the peers. In the unstructured network, peers can leave and join at any time. This can be exploited by an adversary that can join and see the messages floating in the network and can further do source spoofing, reordering or injecting of messages. 

Blockchain can also use structured P2P network where nodes are organized in a specific topology and thus finding any resource/information becomes easier. In this structured P2P network, an identifier is assigned to each node to route the messages in a more accessible way. Each node also maintains a routing table. A structured P2P network maintains a distributed hash table (DHT) where (key, value) pairs are stored corresponding to the peers which help in the resource discovery. Ethereum has started the adoption of structured P2P network by using Kademlia protocol~\cite{Kademlia}. However, most of the blockchain networks are unstructured, and moreover, if the blockchain is public where no restriction to join or leave the network is enforced, then many possible attacks can happen. Thus, the security of blockchain depends heavily on the network architecture. A propagation delay or a synchronization problem in a P2P network can affect the consensus protocol of blockchain, leading to a non-consistent global view in blockchain. In addition to these problems, an adversary can cause several attacks in a P2P network, where few of the main attacks are as following:

\begin{itemize}
    \item Netsplit (Eclipse) attack: An adversary monopolizes all of the connections of a node and splits that node from the entire network. Further, the node  cannot participate in consensus or validation protocol and this causes inconsistency in the network~\cite{Netsplit}.
    \item Routing attack: A set of participants are isolated from the blockchain network by the  adversary and thus the block propagation is  delayed in the network~\cite{Routing}.
    \item Distributed Denial-Of-Service (DDOS) attack: An adversary exhausts the network resources and targets honest nodes so that honest nodes do not get the services or information which they are supposed to receive~\cite{DDOS, DDOS_Bitcoin}. 
\end{itemize}

\begin{table*}[htbp]
    \centering
    
     \resizebox{\textwidth}{!}{
    \begin{tabular}{|l|l|c|c|l|c|}
         
    \hline
    \textbf{Blockchain Type} & \textbf{Application Domain} & \textbf{Anonymity} & \textbf{Scalability} & \textbf{Challenges} & \textbf{References} \\[3ex]
      \hline
      
      Permissionless Public & Decentralized P2P Networks & High & Low & \makecell{Privacy, \\ Scalability} & \makecell{Bitcoin~\cite{Nakamoto_bitcoin}, Zerocash~\cite{Zerocash},\\ Monero~\cite{Monero}} \\[3ex]
      \hline
      
      Permissioned Public & Decentralized Organizations & High & Moderate & \makecell{Privacy, \\ Centralization} & Ripple~\cite{Ripple}, EOS~\cite{EOS} \\[3ex]
      \hline
      
      Permissionless Private & Intra-Organization Networks & Moderate & Moderate & \makecell{Consensus, \\Scalability} & LTO~\cite{LTO} \\[3ex]
      \hline
      
      Permissioned Private & Organizational restricted ledgers & Low & High & \makecell{Consensus,\\ Centralization} & \makecell{Hyperledger fabric~\cite{Hyperledger},\\ Monax~\cite{Monax}, Multichain~\cite{MultiChain}}\\[3ex]
      \hline

     \end{tabular}
     }
    \caption{Blockchain classification.}
    \label{BC-types}
\end{table*}

\subsection{Types of Blockchain} 
Blockchains can be classified depending on the implementation design, administration rules, data availability, and access privileges. From an academic point of view, they have been classified as "public" and "private". While from the administrative point of view, they are described as "permissioned" and "permissionless". Nevertheless, these terms are used interchangeably in most of the blockchain studies and applications in industries, which is not the correct way to use these terms. Even though the classification of blockchains is not very clearly specified in the literature, we can still classify blockchains by coupling public, private, permissioned and permissionless. 

\begin{enumerate}
    \item \textit{Permissionless Public:} In this type of blockchain, anyone can join or leave the network at any time and participate in consensus as well to maintain the ledger. Everyone also has read and write access to the blockchain. Thus, it provides minimum trust among the participants, but it still achieves maximum transparency. Most of the cryptocurrencies and blockchain platforms are permissionless public, e.g., Bitcoin~\cite{Nakamoto_bitcoin}, Zerocash~\cite{Zerocash} and Monero~\cite{Maxwell}.
    
    \item \textit{Permissioned Public:} This type of blockchain allows everyone to read the blockchain state and data, but in order to write the data and take part in consensus, there are permissions/privileges associated with the participants provided by the network administrator which in a certain way makes the system not fully decentralized. In this type of blockchain once a participant has some privileges, based on that it can become a validator as well. Examples for permissioned public blockchain are Ripple~\cite{Ripple}, EOS~\cite{EOS} and the newest Libra~\cite{Libra}.
    
    \item \textit{Permissionless Private:} This type of a blockchain allows organizations to collaborate without the need of sharing information publicly. Being permissionless, allows anyone to join or leave the blockchain at any time, which is also acknowledged by other nodes as well. The smart contracts on these networks also define who is allowed to read the contract and the related data, not only just who is allowed to perform the actions.
    Some permissionless private blockchains use Federated byzantine agreement as a consensus protocol. LTO~\cite{LTO} network is an example of a permissionless private blockchain which creates "live contract" on the network.
    
    \item \textit{Permissioned Private:} These blockchains are mostly used in organizations where data/
    information is stored in the blockchain with permissioned access control by members of the organization. The membership in the network is provided by the network administrator or some membership authority. Read and write access to the data is also provided by the network administrator. Hyperledger fabric~\cite{Hyperledger}, Monax~\cite{Monax}, Multichain~\cite{MultiChain} are examples of permissioned private blockchains.
    
\end{enumerate}

Table~\ref{BC-types} proffers a clear picture of the classification of blockchains with associated advantages, challenges and application domains. However, in general, permissionless public blockchains are commonly referred to as public blockchains and permissioned private blockchains are referred to as fully private blockchains. A combination of permissioned public and permissionless private makes "consortium blockchain" which is also called a federated blockchain. A consortium blockchain is neither completely public nor completely private, and it makes blockchain as partially decentralized. In consortium blockchain, the consensus is reached by a selected group of participants. Nowadays most of the organizations have embraced consortium blockchains for their blockchain enabled solutions.

\section{Challenges in Blockchain} \label{challenges}
Blockchain as an emerging technology comes with many challenges. In order to solve these challenges, various solutions have been proposed and implemented in the blockchain. The proliferation of cryptocurrencies across multiple payment systems brings many risks in social, economic and technical terms. Blockchain encounters many challenges due to network architecture, underlying consensus protocol and applied cryptographic primitives. Some of these major challenges are security and privacy associated with blockchain, scalability of blockchain, and resource consumption (computational power, memory, network bandwidth). An insightful analysis on the research perspectives and challenges for bitcoin and other cryptocurrencies~\cite{JosephSOK} has been presented in the past and gives a nice overview of scalability, security, privacy and consensus of cryptocurrencies. 

We can summarize our discussion in Section \ref{ConsensusMechanisms}, in a form of generic research problems and research challenges in the area of blockchain consensus mechanisms as follows. Construct a new blockchain consensus mechanism that is better than the existing ones from the following perspectives:
\begin{enumerate}
    \item Less energy consumption;
    \item More efficient consensus achievements;
    \item Better security than the existing consensus mechanisms.
\end{enumerate}

However, further in the paper when we identify a more concrete and focused research challenge, we formulate it in a form of a Research Problem. For example, from the discussion given in the \ref{Hash Function} we can formulate the following:

\vspace{1.0mm}
\begin{researchproblem}\label{ResearchProblemASICvsGPUvsCPU}
Construct sustainable blockchain systems that have one of the following properties: 
\begin{enumerate}
    \item They are provably resistant to give mining advantages to ASIC miners as opposite to GPU and CPU miners;
    \item They are provably resistant to give mining advantages to ASIC and GPU miners as opposite to CPU miners.
\end{enumerate} 
\end{researchproblem}
\vspace{1.0mm}

\begin{table*}[htbp]
\centering

\resizebox{\textwidth}{!}{
\begin{tabular}{|c|c|c|c|c|c|}
\cline{2-6} \multicolumn{1}{c|}{} &
\textbf{Confidentiality} & \textbf{Integrity} & \textbf{Availability} & \textbf{Data Privacy} & \textbf{Anonymity}\\[3ex]
\hline
\textbf{Smart Contract} & Encryption & MAC & -- & \makecell{Data Privacy \\Preserving Computation} & \makecell{Identity Privacy \\Preserving Computation} \\[3ex]
\hline
\textbf{Transaction} & -- & \multirow{2}{*}{Signature Scheme} & \makecell{Access Structure\\ of Transactions} & \makecell{Zero-Knowledge Proofs, \\ Mixing Techniques} & \makecell{Zero-Knowledge Proofs}\\[3ex]
\cline{1-2}\cline{4-6}
\textbf{Consensus} & -- &  & Consensus & Access Control & \makecell{Blind or Ring \\ Signature}\\[3ex]
\hline
\textbf{Network} & \multirow{2}{*}{Encryption} & \multirow{2}{*}{MAC} & Protocols e.g.\ Gossip & -- & IP Anonymity e.g.\ TOR\\[3ex]
\cline{1-1}\cline{4-6}
\textbf{Database} & & & Access Control & Access Control & --\\[3ex]
\hline
\end{tabular}}
\caption{Layered architecture of blockchain.}
\label{Layered-Blockchain}
\end{table*}

If we observe the blockchain as a layered architecture, we can identify the challenges that occur in each layer. Table~\ref{Layered-Blockchain} shows blockchain as a stack of five layers. These five layers serve the following purposes: 
\begin{itemize}
    \item \textbf{Smart contract layer} processes contract data and send the result data to the transaction layer.
    \item \textbf{Transaction layer} creates the transactions and sends those to consensus layer.
    \item \textbf{Consensus layer} runs the consensus algorithm and adds the transactions to the block.
    \item \textbf{Network layer} deals with all P2P communication among blockchain nodes.
    \item \textbf{Database layer} stores the blockchain data in
    a respective database used by respective blockchain platform.
    
\end{itemize}

Table~\ref{Layered-Blockchain} gives a glimpse of blockchain layered architecture and also mentions some of the cryptographic techniques to achieve properties like security and privacy. In Table~\ref{Layered-Blockchain}, the first column defines the layers of blockchain, and the first row illustrates the properties which can be accomplished in the different layer using different cryptographic techniques. Thus to understand, each cell corresponds to the deployed cryptographic method to attain the property in the corresponding column in the respective blockchain layer (corresponding row). For example, encryption can be used to achieve confidentiality in smart contract layer, Message Authentication Code (MAC) can be used to achieve integrity in the network layer of blockchain.
Table~\ref{Layered-Blockchain} names few of the techniques used in the blockchain but there are more available cryptographic techniques which can be employed in blockchain. ``--" in Table ~\ref{Layered-Blockchain} represents that the corresponding property for the corresponding layer does not make much sense.
Some of the significant challenges of blockchain are as follows.

\subsection{Security and Privacy} For any blockchain, a key evaluation parameter is how well the security and privacy conditions meet the requirement of the blockchain. Analyzing the security and privacy issues of blockchain is a broad research area, and some studies have been conducted in this area. Here we do not cover those details, instead we only define these terms. Security is defined as three components: confidentiality, integrity, and availability. In a generic context, (i) confidentiality is a set of rules that limits access to information, (ii) integrity is the assurance that the information is trustworthy and accurate, and (iii) availability is a guarantee of reliable access to the information by authorized people. However, in case of blockchain, the term \textit{Information} used in the above context can have multiple meanings such as data in the database, smart contract data or transactions. Privacy can be defined as data privacy and user privacy (anonymity). Table~\ref{Layered-Blockchain} includes some cryptographic mechanisms for achieving security and privacy of information subjected to different blockchain layers. 

In the light of recent increased number of incidents with the security of the different layers of blockchain platforms and the theft of millions of dollars worth cryptocurrencies, we formulate the following research problem.

\vspace{1.0mm}
\begin{researchproblem}\label{ResearchProblemS&P}
Construct a penetration testing tool irrespective of the blockchain platform to test the security and privacy requirements for each layer of any blockchain platform.
\end{researchproblem}

\subsection{Scalability Issues} The size\footnote{\url{https://bitinfocharts.com} gives most of the statistics (including size) of popular cryptocurrencies.} of blockchain is continuously growing, and scalability is becoming a big problem in the blockchain domain. Scalability depends on the underlying consensus, network synchronization and architecture. To scale the blockchain, the computational power and the bandwidth capabilities should be high for each node in the blockchain, which is practically infeasible. Most of the current blockchains grant limited scalability.

One proposal how to address the scalability problems of the blockchain ledger is so called: "SPV, Simplified Payment Verification" \cite{SVP}. It verifies if particular transactions are valid but without downloading the entire ledger. This method is used by some wallet and lightweight Bitcoin clients, and its security was first analyzed in \cite{skudnov2012bitcoin}. Another proposal to achieve high scalability is to use erasure codes in blockchain by encoding validated blocks into small number of coded blocks. A recent work~\cite{Fountain-codes} proposes the use of fountain codes (a class of erasure codes) to reduce the storage cost of blockchain by the order of magnitude and hence achieving high scalability. Applying other types of erasure codes for distributed storage, such as regenerating codes \cite{5550492,8025778}, locally repairable codes \cite{6259860,7593121} or a combination of both types of codes \cite{6846301,8328506}, may reduce even further the storage and communication costs.

Another issue in connection with the scalability is the issue of the interoperability. Namely, it is a fact that the number of different public ledgers is increasing rapidly. While some sort of a rudimentary interoperability has been implemented in cryptocurrencies exchange platforms \cite{white2015market}, the risks and insecurities with these platforms are vast and well documented \cite{mclannahan2014bitcoin}.
\vspace{1.0mm}
\begin{researchproblem}\label{ResearchProblemSmallerPublickLedger}
Construct a new blockchain mechanism that periodically prunes its distributed ledger (reduces its size), producing a fresh but equivalent ledger, while provably keeping correct state of all assets that are subject of the ledger transactions.
\end{researchproblem}
\begin{researchproblem}\label{ResearchProblemInteroperability}
Construct secure protocols for blockchain interoperability.
\end{researchproblem}

A recent reference~\cite{Vitalik-Scalability} strongly supports our research problem~\ref{ResearchProblemSmallerPublickLedger} since it admits that Ethereum blockchain is almost full now and hence the scalability is a big bottleneck. 

\subsection{Forking} 
\begin{figure}[h]
    \centering
    \includegraphics[width = 0.90\textwidth]{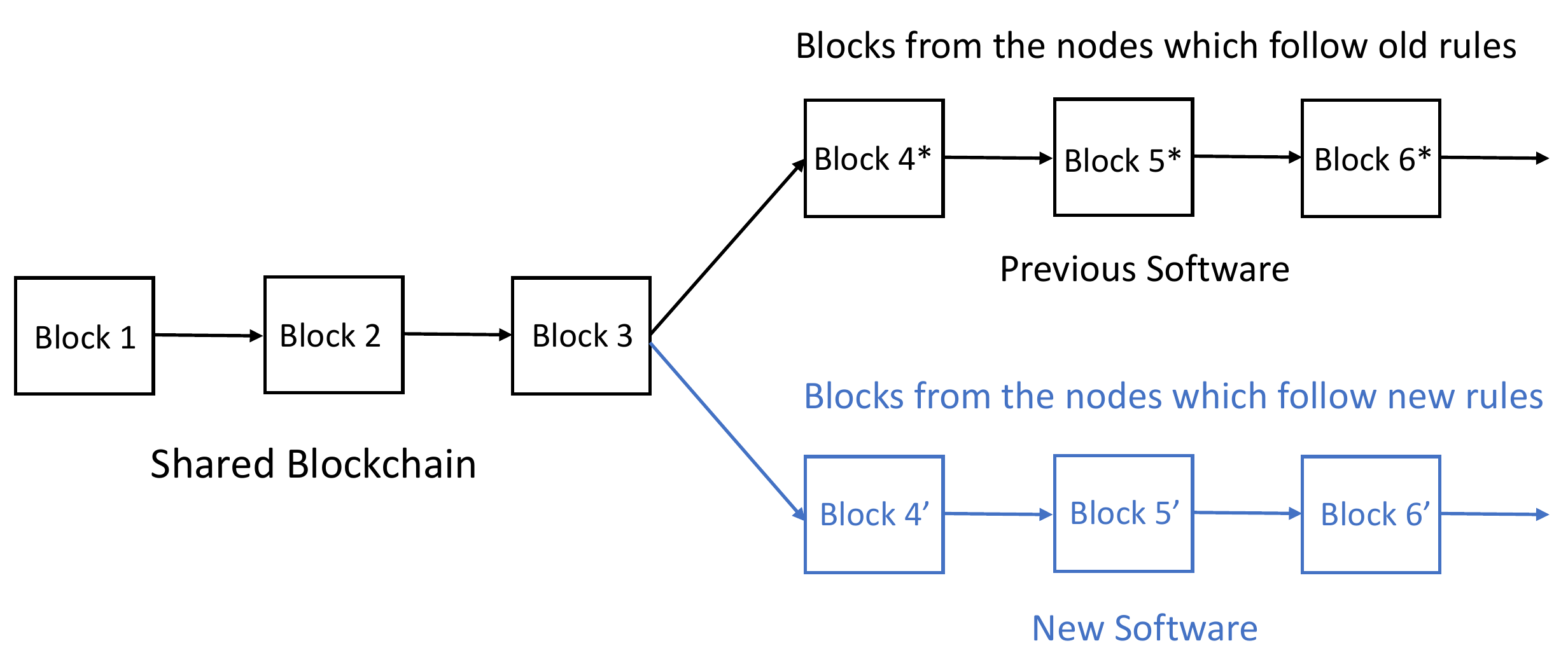}
    \caption{Blockchain forking.}
    \label{Fork}
\end{figure}

A blockchain fork is essentially caused when two miners find a block at almost the same time due to a software update or versioning. In a blockchain network, each device or computer is considered as "a full node" which runs software to keep the blockchain secure by verifying the ledger. The software is updated to adjust some parameters and to install new features in the blockchain. This updated software may not be compatible with the old software. Consequently, the old nodes which have not updated their software and the new nodes which have performed a software update can cause a fork in the blockchain when they create new blocks. There are two types of forks: one which is not compatible with previous software version, called a hard fork, and another one which is compatible with the previous version (backward-compatible), called a soft fork. A hard fork happens when there is a significant change in the software such as change of block parameters or change of consensus mechanism. In the case of Ethereum, a hard fork will occur when it will migrate from Proof of Work to Proof of Stake. One example of a soft fork is Segregated Witness (SegWit) which was implemented in Bitcoin by changing the transaction format. Recently, privacy coin Beam~\cite{Beam} (an implementation of Mimblewimble privacy protocol) conducted its first hard fork away from ASICS. Figure~\ref{Fork} depicts a blockchain forking scenario where the correct chain can be any of these two forked chains depending on the case of the hard or soft fork.
\vspace{1.0mm}
\begin{researchproblem}\label{ResearchProblemForkning}
Construct Forking-free consensus mechanism for permissionless public blockchain.
\end{researchproblem}

\subsection{Throughput} It is a measure of the number of blocks appended in blockchain per second which effectively means the number of transactions processed per second. Throughput depends on many factors such as underlying consensus algorithm, number of nodes participating in consensus, network structure, node behavior, block parameters and the complexity of the contract (in case of smart contract supported blockchains). The complexity of a smart contract depends on whether the programming language of the blockchain is turing-complete or not. However, regarding turing-completeness of blockchains~\cite{Turing-Completeness}, there is always a division between the blockchain community. 
Considering these primary factors, attaining high throughput is a bit hard in blockchain. However, to achieve high throughput, the size of the transaction can be reduced by excluding some information from the transaction. The throughput can be increased by increasing the block size and the bandwidth of the network till a certain level.

The number of transactions per second was recognized as a serious problem in Bitcoin network. While in the peak holiday period Visa and MasterCard can handle up to 50,000 transactions per second worldwide, the Bitcoin network can handle just 7 transactions. One proposal how to address this scalability issue is the "The Bitcoin Lightning Network" \cite{poon2016bitcoin}. It is a network that handles instantly the Bitcoin transactions off the main ledger. It establishes a network of micropayment channels that addresses the malleability by using Bitcoin multi-signatures 2-of-2. Special nodes are needed for these micropayment networks and as of June 2019, there were around 4,500 nodes. The first financial transaction via the Lightning network was reported in January 2018. 
Litecoin decided to follow the Bitcoin Lightning network, and as of March 2019 there were more than 1000 registered nodes that handle the micropayments for that alternative cryptocurrency. Many other solutions were proposed to solve the scalability issue, similar to the Lightning off-chain computation and off-chain state channels, such as Sharding~\cite{Sharding}, Plasma~\cite{Plasma}, Liquid~\cite{back2014enabling} and the recent Channel Factories~\cite{burchert2018scalable}.

As the Lightning network is gaining in popularity, new research challenges emerge as depicted in \cite{PIR}, and here we rephrase one of their research challenges:
\vspace{1.0mm}
\begin{researchproblem}[~\cite{PIR}]\label{ResearchProblemLightningNetwork}
Develop scalable protocols that will perform multi-hop payment-channel and path-based transactions with strong privacy guarantees even against an adversary that has network-level control.
\end{researchproblem}

Addressing Problem~\ref{ResearchProblemLightningNetwork}, many works have been done in the past but all those works are mostly compatible with Bitcoin or Ethereum blockchain. Recent works~\cite{AMCU,AMHL} on multi-hop payment channel provide value privacy and security but only for Bitcoin-compatible blockchains. 
Instead of supporting only payments like Lightning network, there are off-chain state channels, like Celer Network~\cite{Celer}, which support general state updates while providing significant improvement in terms of cost and finality. 
\vspace{1.0mm}
\begin{researchproblem}\label{ResearchProblemStateChannel}
Develop fully functional state channel with strong security and privacy guarantee.
\end{researchproblem}

\subsection{Energy Consumption} The mining process of blockchain (e.g., bitcoin mining) consumes a lot of energy. Most of the PoW puzzle based consensus protocols waste a huge amount of energy\footnote{\url{https://digiconomist.net/bitcoin-energy-consumption} depicts Bitcoin energy consumption index charts in TWh per year. It also shows the energy consumption per country.}. Many alternative consensus algorithms are introduced which use less energy than Bitcoin's PoW such as Proof of Stake~\cite{Proof-of-stake}, Equihash~\cite{Equihash}, and PBFT~\cite{PBFT}. Energy is also consumed during communication over the network. Some cryptographic mechanisms also consume high energy so the selection of a proper cryptographic mechanism should be based not only on the memory requirement and the computational load but also on the amount of energy consumption. The use of blockchain should be energy efficient and to fulfill that 1) PoS-like consensus should be used and 2) proper energy management techniques should be utilized, for example in the case of Internet-of-Things (IoT).

\subsection{Infrastructure Dependencies} The blockchain infrastructure is built with several elements of network protocols, cryptographic concepts, and mining hardware. All these elements depend on each other in some sense. If we look into the layered architecture of blockchain in Table~\ref{Layered-Blockchain}, each layer is dependent on its upper and lower layers for some input/output. Thus, there are many infrastructure dependencies in blockchain. For instance, the data from the smart contract layer is an input to the transaction layer that outputs actual transactions; the data from the consensus layer results in an input to the network layer through a communication protocol; and the data from the network layer data is sent to the database through database storage management. These dependencies must be taken into account while building a comprehensive blockchain framework for any use case; otherwise, some of the blockchain functionalities will not be fulfilled. 

From the blockchain infrastructure perspective, we have to mention here one evolving and enabling technology that will be very important in the next decade: 5G. 5G will connect hundreds of billions of IoT devices, but that vast number of devices can be governed securely only by strong decentralized mechanisms offered by the blockchain technologies~\cite{Kshetri2018,HillDeweyPlasencia2018}. We formulate this debate as the following.
\vspace{1.0mm}
\begin{researchproblem}\label{ResearchProblemIoT5G}
Construct efficient, scalable, inexpensive and sustainable blockchain systems capable to handle and securely manage up to billions of IoT devices connected via the 5G network infrastructure.
\end{researchproblem}

\section{Overview of used cryptographic concepts in blockchain}\label{UsedCryptoConcept}
From the cryptographic point of view, many of the cryptographic techniques have already been exhibited and heavily employed in various blockchain platforms and blockchain use-cases~\cite{CryptoPrimitives}. As the spectrum of the cryptographic concepts is vast, there is always a scope to dig out some of the existing cryptographic schemes and use them in blockchain services.

In Table \ref{Techniques-table} we give a comprehensive summary of all cryptographic concepts that we will cover in this and in the next Section. It serves as a handy overview and quick reference table for our systematization of the cryptographic knowledge used in blockchain.

Following are some of the cryptographic concepts which have already been well analyzed and implemented in blockchain.

\begin{table*}[htbp]
\centering

\resizebox{\textwidth}{!}{
\begin{tabular}{|l|l|l|}
\hline
\textbf{Cryptographic Concept} & \textbf{Properties} & \textbf{Instantiation (Reference)}\\
\hline
Access Control & Data privacy & Hyperledger Fabric~\cite{Hyperledger}, FairAccess~\cite{FairAccess}, ~\cite{IoTAC}\\
\hline
Accumulator & Provides Membership Proofs, Anonymity  & Batching Techniques for Accumulators in Blockchain~\cite{Accumulator}\\
\hline
Aggregate Signature & Fast Signature Verification & Tested in Bitcoin~\cite{AggSign}\\
\hline
Commitment Scheme & Non-Repudiation & Used in Bullteproof~\cite{Bunz} and in Monero~\cite{Monero, Maxwell}\\
\hline
Decentralised Authorization & Data Privacy & BlendCAC~\cite{BlendCAC}, WAVE~\cite{WAVE}\\
\hline
Encryption Scheme & Confidentiality and Anonymity & Kadena~\cite{Kadena}, Hyperledger Fabric~\cite{Hyperledger}, Tendermint~\cite{tendermint} \\
\hline
Identity Based Encryption & No Public Key Distribution Infrastructure & BAVP~\cite{BAVP}, BLIC~\cite{BLIC}\\
\hline
Incremental Cryptography &  Efficiency Improvement & Kadena~\cite{Kadena}\\
\hline
Lightweight Cryptography & Fast, less Memory/Energy Consumption & LSB~\cite{LSB}, EVCE~\cite{EVCE}\\
\hline
Obfuscation & Privacy & Tested in Bitcoin~\cite{Obfuscation}\\
\hline
Oblivious RAM & Confidentiality and Integrity & Solidus~\cite{Solidus}, EVORAM~\cite{EVORAM}\\
\hline
Oblivious Transfer & Data Privacy & Searchain~\cite{Searchain}, ~\cite{APDB}\\
\hline
Post-Quantum Cryptography & Quantum Resistant & Post-Quantum Blockchain~\cite{PQB}, ~\cite{PQBCurrency}, ~\cite{QABitcoin}\\
\hline
Private Information Retrieval & Data Privacy & Private Blockchain Queries from PIR~\cite{PIR}\\
\hline
Proof of Retrievability & Cloud Data Recovery & Permacoin~\cite{Permacoin}, Retricoin~\cite{Retricoin}, Storj~\cite{Storj}\\
\hline
Secret Sharing & Data privacy & SHARVOT~\cite{SHARVOT}, Wanchain~\cite{Wanchain}\\
\hline
Secure Multiparty Computation & Privacy of Peers and Smart Contract  & Enigma~\cite{Enigma}, Hawk~\cite{Hawk}, Wanchain~\cite{Wanchain}\\
\hline
Signature Scheme & Integrity and Authentication  &  In Every Blockchain e.g. Multichain~\cite{MultiChain}, CryptoNote~\cite{CryptoNote}\\
\hline
Verifiable Delay Function & Less Parallelism, Fast Verification & Chia Network~\cite{Chia}\\
\hline
Verifiable Random Function & Verifiable Pseudorandom Output & Algorand~\cite{Algorand}, Ouroboros Praos~\cite{Ouroboros_Praos}, Dfinity~\cite{Dfinity}\\
\hline
White-Box Cryptography & Data Privacy & Runtime Self-Protection in Blockchain Ledger~\cite{White-box}\\
\hline
Zero-Knowledge Proof & User and Data privacy & Zerocoin~\cite{Zerocoin}, Zerocash~\cite{Zerocash}\\
\hline
\end{tabular}
}
\caption{Summary of Cryptographic Concepts in Blockchain.}
\label{Techniques-table}
\end{table*}


    \subsection{Signature Scheme}
   A standard digital signature is a mathematical scheme based on public-key cryptography that aims to produce short codes called signatures of digital messages by the use of a private key, and where those signatures are verifiable by the use of the corresponding public key. In this context, digital signatures guard against tampering and forgeries in digital messages.

   \begin{figure}[htbp]
        \centering
        \includegraphics[width = 0.55\textwidth]{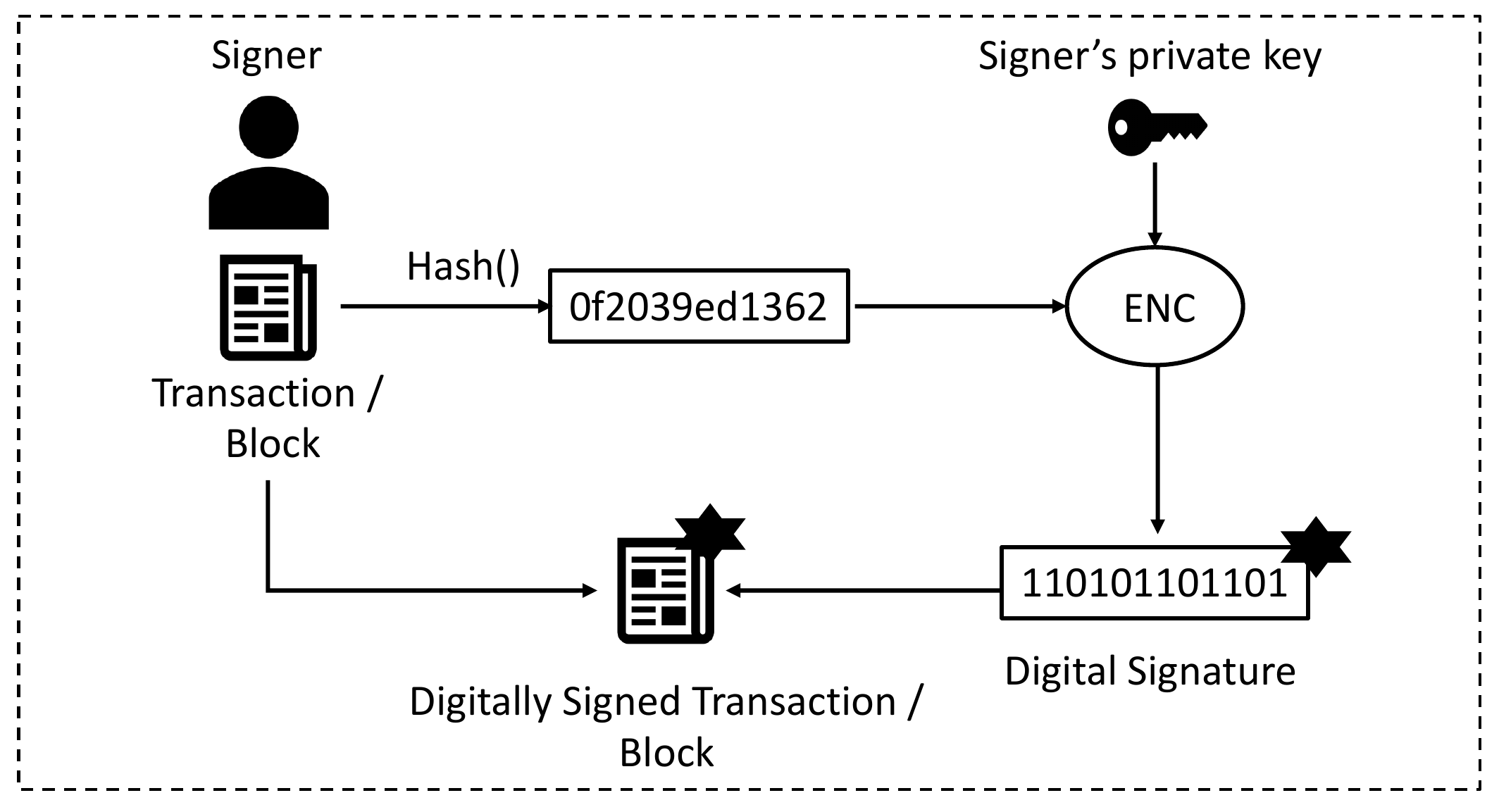}
        \caption{Signing process of blockchain transaction/block}
        \label{Signature}
    \end{figure}
    
    \begin{figure}[htbp]
        \centering
        \includegraphics[width = 0.55\textwidth]{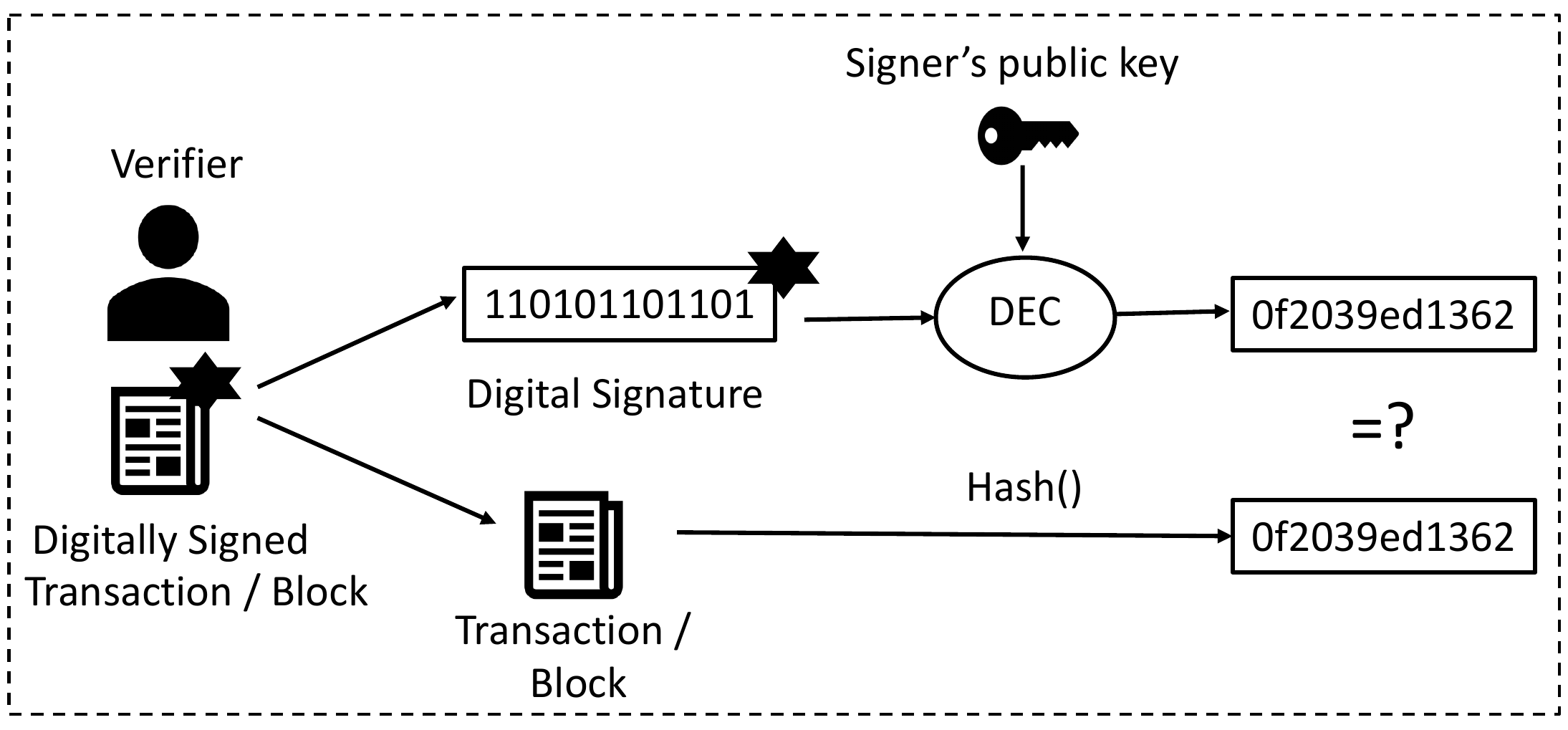}
        \caption{Verification of digitally signed transaction/block}
        \label{Signature-Verification}
    \end{figure}
   
    A signature scheme is used in blockchain to sign the transaction, hence, authenticating the intended sender and providing transaction integrity as well as non-repudiation of the sender. Many of the signature schemes are widely accepted to employ integrity and anonymity in blockchain. The digital signature is one of the most important cryptographic primitives that makes blockchain to be publicly verifiable and with achievable consensus. Signature schemes are used in almost every blockchain. Figure~\ref{Signature} represents a general example about how a blockchain user (signer) creates a digitally signed transaction or block using his private key. Moreover, figure~\ref{Signature-Verification} shows how other blockchain nodes (verifier) verify whether the signature on the transaction or block is valid or not using the signer's public key. Blockchain applies different signature schemes to provide additional features like privacy, anonymity, and unlinkability. Signature scheme can also be applied to generate constant size signature using signature aggregation. Schnorr Signatures are a form of signature aggregation and it has been used in Bitcoin instead of P2SH~\cite{P2SH} for scalability~\cite{Schnorr}. Some of the signature schemes applied in blockchain are:
    
    \begin{enumerate}
    \item Multi-Signature: In a multi-signature scheme, a group of users signs a single message. In a blockchain network, when a transaction requires a signature from a group of participants, it might be advantageous to use a multi-signature scheme. Blockchain platforms such as Openchain~\cite{Openchain} and MultiChain~\cite{MultiChain} support $M-$of$-N$ multi-signature scheme which reduces the risk of theft by tolerating compromise of up to $M$-1 cryptographic keys. Boneh et al. also designed compact multi-signatures for smaller blockchains~\cite{Multi-signature}.
    \item Blind Signature: In this scheme~\cite{Chaum84}, signatures are employed in privacy-related protocols where the signer and the message authors (transaction in case of blockchain) are different parties. Blind signatures are used to provide unlinkability and anonymity of the transaction. In a blockchain setup, a blind signature might be helpful to provide anonymity and unlinkability where the transacting party and the signing party are different. Blind signatures have been used in BlindCoin~\cite{BlindCoin} distributed mixing network to provide the unlinkability of transactions. Blind signatures are also tested in Bitcoin to provide the anonymity for the Bitcoin on-chain and off-chain transactions~\cite{Blind_bitcoin}. 
     \item Ring Signature: This scheme~\cite{Zhang02} uses a protocol where a signature is created on a message by any member of a group on behalf of the group while preserving the identity of the individual signer of the signature. Ring signatures are used to achieve anonymity of the signing party in the blockchain network. CryptoNote~\cite{CryptoNote} technology uses a ring signature scheme to create untraceable payments in the cryptocurrencies. A trustless tumbling platform~\cite{Mobius18} also uses ring signature for anonymity.
     \item Threshold Signature: This signature scheme is a $(t, n)$ threshold signature where $n$ parties receive a share of the secret key to create the signature and $t$ out of $n$ parties create a signature over any message. As the parties directly construct the signature from the shares, the key is never revealed in the entire scheme. Threshold signature can be helpful to provide anonymity in the blockchain network. CoinParty~\cite{CoinParty} uses a threshold signature scheme for multi-party mixing of Bitcoins. A recent work about coin mixer, ShareLock~\cite{Sharelock}, uses threshold ECDSA (Elliptic Curve Digital Signature Algorithm~\cite{ECDSA}) to provide privacy-enhancing solution for cryptocurrencies. However threshold ECDSA signatures are complex due to the intricacies of the signing algorithm. Other signature schemes, such as EdDSA  (Edwards-curve Digital Signature Algorithm~\cite{EdDSA}) using the Edwards25519 curve, are efficient threshold signatures. Libra~\cite{Libra} blockchain applies this EdDSA during new account address generation.
\end{enumerate}  

While digital signatures produced with the keys used in Public Key Infrastructure (PKI) are well legally regulated and can be used in different types of legal disputes, it is a big challenge how to achieve similar regulations with all types of digital signatures used in the existing blockchain solutions. Additionally, in the physical world if an asset is stolen (for example an expensive car, or an expensive watch), it can be traced back to its legal owner.

\vspace{1.0mm}
\begin{researchproblem}\label{ResearchProblemPKIBlockchain}
Develop security protocols that merge the existing standardized and legalized PKI systems with some of the the developed blockchain systems.
\end{researchproblem}
\begin{researchproblem}\label{ResearchProblemStolenAssets}
Design an anti-theft blockchain system, i.e., a system that guarantees a return of stolen assets back to their legitimate owners.
\end{researchproblem}

 Regarding Research Problem~\ref{ResearchProblemStolenAssets}, recently the Vault proposal was re-launched. Its purpose is to shield the bitcoin wallet from theft without the need for hard forking~\cite{Vault}. However, for other blockchain systems, no such proposal or solution exists.
 
\subsection{Zero-Knowledge Proofs} In Zero-knowledge proofs~\cite{Goldreich94}, two parties, a prover and a verifier, participate. First, the prover asserts some statement and proves its validity to the verifier without revealing any other information except the statement. Thus, a zero-knowledge proof proves the statement as `transfer of an asset is valid' without revealing anything about the asset. Zero-knowledge protocols are extremely useful cryptographic protocols for achieving secrecy in the applications. They can be used to provide the confidentiality of an asset (transaction data) in the blockchain while keeping the asset in the blockchain. Some of the public blockchains use zero-knowledge proofs such as Zerocoin~\cite{Zerocoin} or Zerocash~\cite{Zerocash} for untraceable and unlinkable transactions. Zerocoin is a decentralized mix and extension to Bitcoin for providing anonymity and unlinkability of transactions by applying zero-knowledge proofs. In Zerocoin protocol, a user who has Bitcoins can generate an equal value of Zerocoins without the need of any third party mixing set. A user can spend his/her Bitcoin by 1) producing a secure commitment (i.e., Zerocoin), 2) recording it in the blockchain, and 3) broadcasting a transaction and a zero-knowledge proof for the respective Zerocoin. Hence, other users can validate the Zerocoin recorded in the blockchain and verify the transaction along with the proof. Here zero-knowledge proof protects the linking of Zerocoin to a user, 
    yet Zerocoin is a costly protocol due to its high complexity and large proof size. 
    
    \begin{figure}[htbp]
        \centering
        \includegraphics[width = 0.9\textwidth]{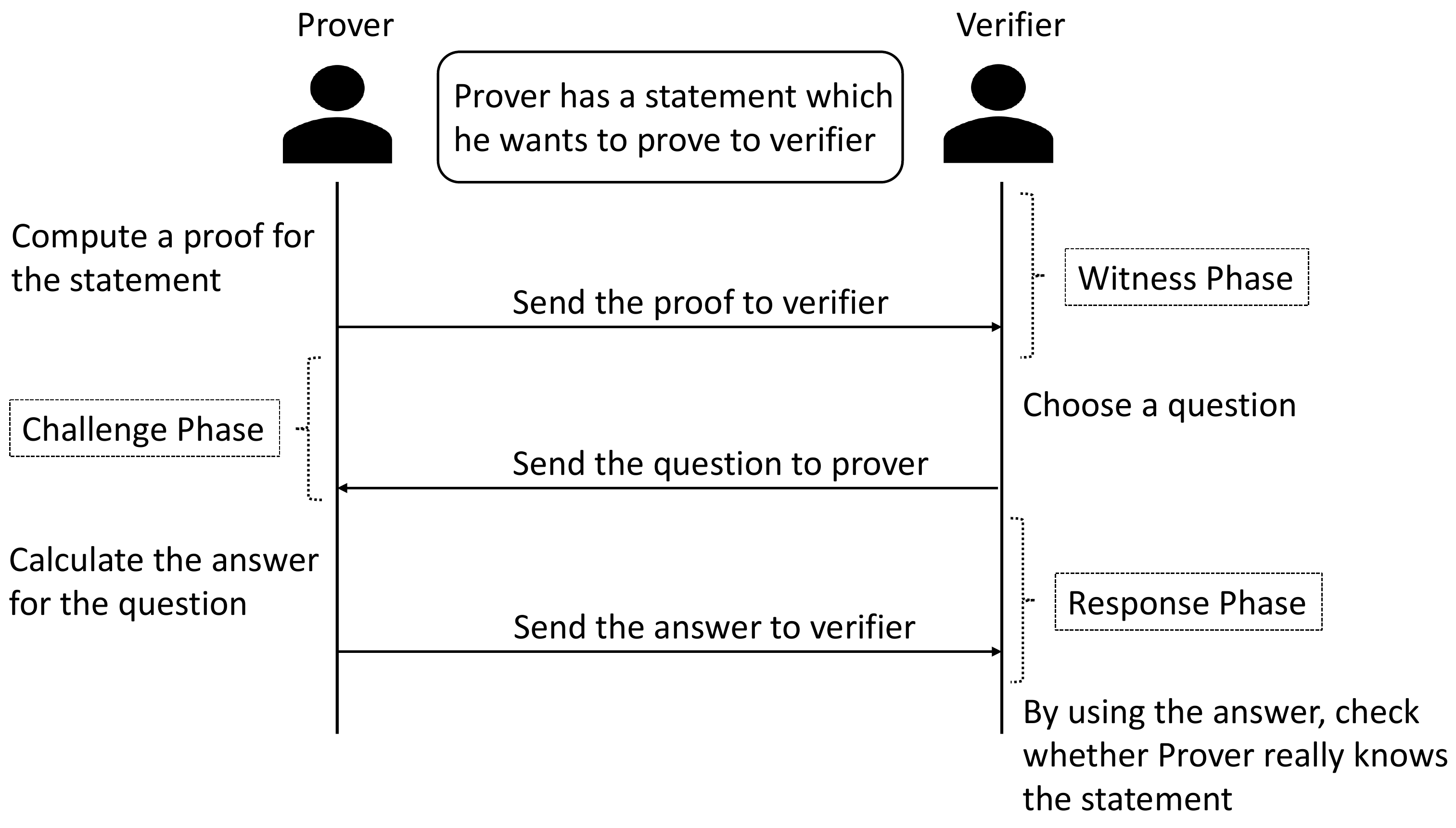}
        \caption{An interactive zero-knowledge protocol.}
        \label{Zero-knowledge}
    \end{figure}

    To reduce the complexity and the proof size, a variant of zero-knowledge proof known as Zero-Knowledge Succinct Non-Interactive Argument of Knowledge (zk-SNARK)~\cite{zk-SNARK} is used by Zerocash protocol. zk-SNARK hides the information about the amount and the receiver address in a transaction. The main idea of zk-SNARK is `any computational condition can be represented by an arithmetic circuit, which takes some data as input and gives \textit{true} or \textit{false} in response'. zk-SNARK reduces the proof size and the computational effort compared to the basic zero-knowledge proofs. An enterprise-focused version of Ethereum, Quorum blockchain platform~\cite{Quorum} also uses zk-SNARK for transaction privacy and anonymity. Figure~\ref{Zero-knowledge} illustrates an interactive protocol of zero-knowledge where the prover has a statement, and he/she wants to prove that the statement is correct without revealing any information related to the statement. In the interactive protocol, the verifier asks many questions related to the statement and the prover answers these questions in such a way where the prover proves the statement and does not reveal any necessary information.

    \subsection{Access Control} It is a selective restriction on information or resource based on some policy or criteria. These mechanisms~\cite{Sandhu94} can be enforced to put a restriction or access in the blockchain. The access can be a read/write access or an access to participate in a blockchain protocol. There are many different access control mechanisms such as role-based, attribute-based, organizational-based access control which can be used in blockchain. Recent incidents show security breaches and data theft from certain blockchain platforms, which can be tackled and prevented by access control. The privacy of data can be ensured in blockchains by using access control~\cite{IoTAC, FairAccess}. Nowadays, access control techniques are profoundly used in blockchain based medical applications~\cite{MedRec} or blockchains for the insurance industry where the data is sensitive information that must be accessible to only trusted and authorized parties. There are different types of access control mechanisms which can be utilized in blockchain applications; some of these mechanisms are explained to be used in the blockchain.
    
    \begin{enumerate}
    \item Role-based Access Control (RBAC): RBAC is an approach for restricting the system view to the users of the system according to their roles in the system. Thus, it can be applied in a blockchain framework where access is provided according to the user roles. RBAC is used in a blockchain based solution for Healthcare~\cite{RBAC}. A simple example depicted in Figure~\ref{Access-Control} describes the role-based access control in a private healthcare blockchain. Based on the role, each entity in the blockchain system has its own access rights. A Patient can ask for his personal medical data, however only the Doctor associated with the patient can enter or modify the patient's health record in the blockchain. A Research Company on the other hand can ask for patients' data for any disease for research purpose. 
    
    \begin{figure}[htbp]
        \centering
        \includegraphics[width = 0.7\textwidth]{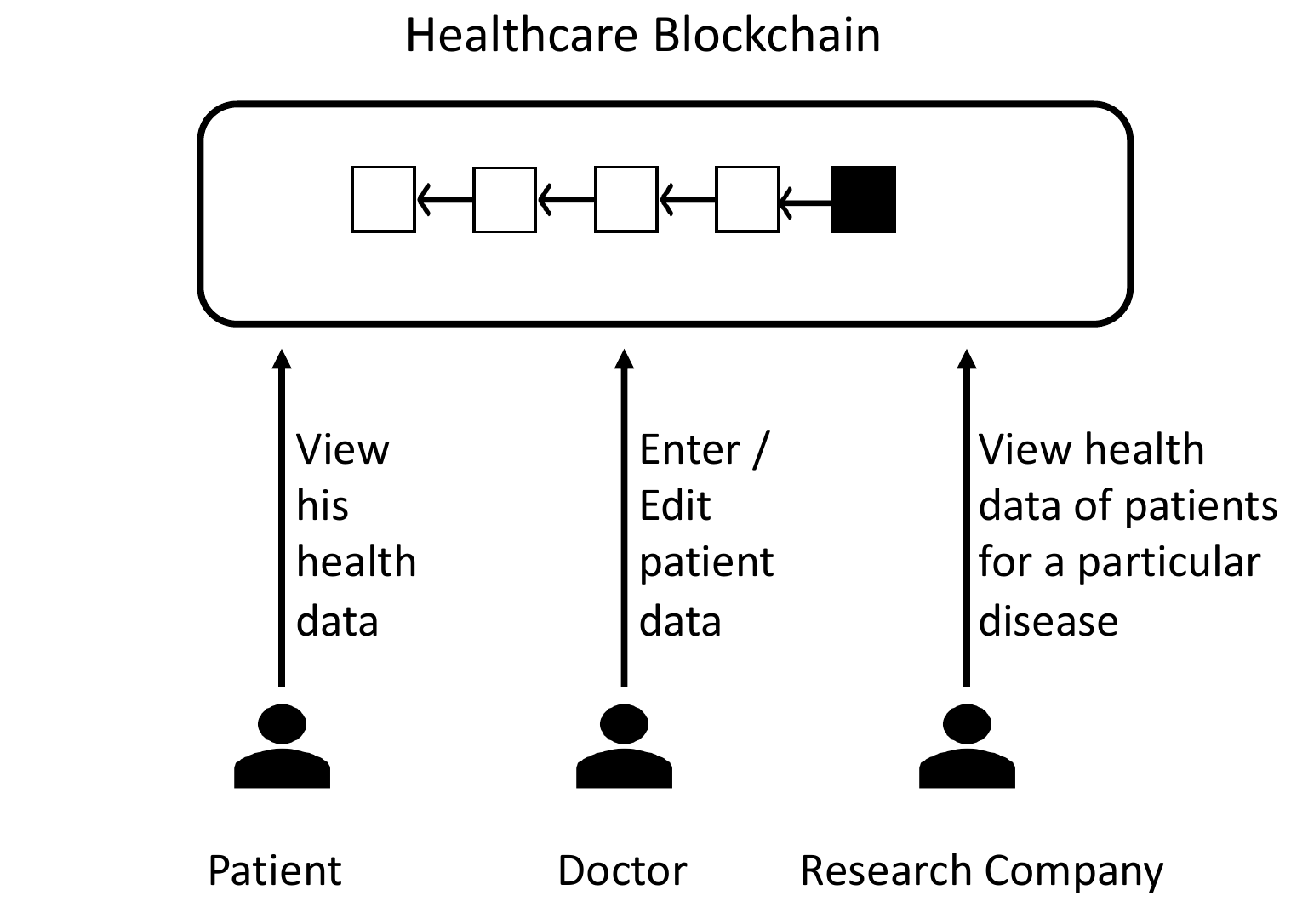}
        \caption{Role-based access control in healthcare blockchain }
        \label{Access-Control}
    \end{figure}

    \item Attribute-based Access Control (ABAC): In ABAC, the access control rules are based on the attribute structure. These attributes can be user specific, environment-specific or object specific. For example, in a blockchain setup for the insurance industry, 'department' could be an attribute through which the access of the blockchain data is restricted, which means the claims handling department would have a different view of the blockchain compared to the audit department. ABAC can be used in a fair access blockchain model~\cite{FairAccess} by keeping attributes in policy.
    
    \item Organization-based Access Control (OrBAC): OrBAC is one of the richest access control models. OrBAC consists of three entities (subject, action, object) which define that some subject has the permission to realize some action on some object. OrBAC has already been used in blockchain for IoT in a fair access blockchain model~\cite{FairAccess} and in dynamic access control model on blockchain~\cite{OrBAC}. 
    
    \end{enumerate}
    Other access control mechanisms such as context-based access control and capability-based access control (proposed in blockchain solutions for autonomous vehicles, smart cities, IoT~\cite{CBAC}) can also be useful for different blockchain solutions.

    \subsection{Encryption Scheme} It is a process of encoding a piece of information by which only authorized parties can access it. It can be used to achieve confidentiality of blockchain data by encrypting it. There are many encryption schemes which can be used in blockchain. Symmetric-key Encryption is used in Hyperledger fabric for confidentiality of smart contract~\cite{Hyperledger} and Blockchain for Smart Home~\cite{IoTBC}. Although searching and computation over an encrypted data is a big challenge, there are many existing techniques which can be used for that purpose. Some of these techniques such as searchable encryption for searching on encrypted data in the cloud is already used in permissioned blockchain~\cite{Tahir}, and for computation over encrypted data, fully-homomorphic encryption and functional encryption can also be utilized in blockchain. Monero cryptocurrency~\cite{Monero} uses (half) additive homomorphic encryption together with range proof techniques, yet supporting only value transactions. 
    
    In order to assure simultaneously confidentiality and authenticity of data, an authenticated encryption can be used in blockchain. In authenticated encryption, two peers establish a connection, they both share their public keys and compute the shared secret which is used as the symmetric key for the authenticated encryption algorithm. The recently finished cryptographic competition CAESAR \cite{bernstein2014caesar} has identified a portfolio of six ciphers for authenticated encryption. So far, as of this writing (June 2019), none of those ciphers has been deployed in some blockchain system.
    
    Broadcast encryption can be used in blockchain to provide the anonymity of blockchain receiver nodes. \cite{BroadcastIoT} gives a proposal to use for Availability and Accountability for IoT by blockchain. It has  as every user in the group receives the encrypted message, although only users with the correct permission or key can decrypt it.

   \subsection{Secure Multi-party Computation (SMPC)}
        Secure Multi-party Computation enables parties to act together in a way that no single party has an access to all of the data, and hence no one can leak any secret information. The main idea of SMPC scheme is to jointly compute a function by parties over their inputs without disclosing their inputs. For example, a group of people can compute the average salary of the group without disclosing their actual individual salaries. The blockchain platform Enigma~\cite{Enigma} leverages the concept of SMPC to achieve strong privacy. In Enigma platform, a blockchain network is combined with SMPC network, where the blockchain network contains the hashes and SMPC network contains the data corresponding to those hashes which split is among different nodes. For each node, the view over SMPC network differs as everyone has a different piece of information. Specifically, each node contains a random piece of data, and no single party ever has access to the entire data.
    
    A blockchain model Hawk~\cite{Hawk} for privacy-preserving smart contracts also specifies the use of SMPC to minimize the trust in the generation of common reference string in SNARK proof used in the model. SMPC can also be exercised for private data storage in a decentralized system, such as Keep~\cite{Keep}. Keep provides a privacy-focused storage solution for Ethereum. In this system, network nodes collaborate to provide secure decentralized data containers, called keeps, which can be accessed from smart contracts on Ethereum. 
    
     \begin{figure}[htbp]
        \centering
        \includegraphics[width = 0.7\textwidth]{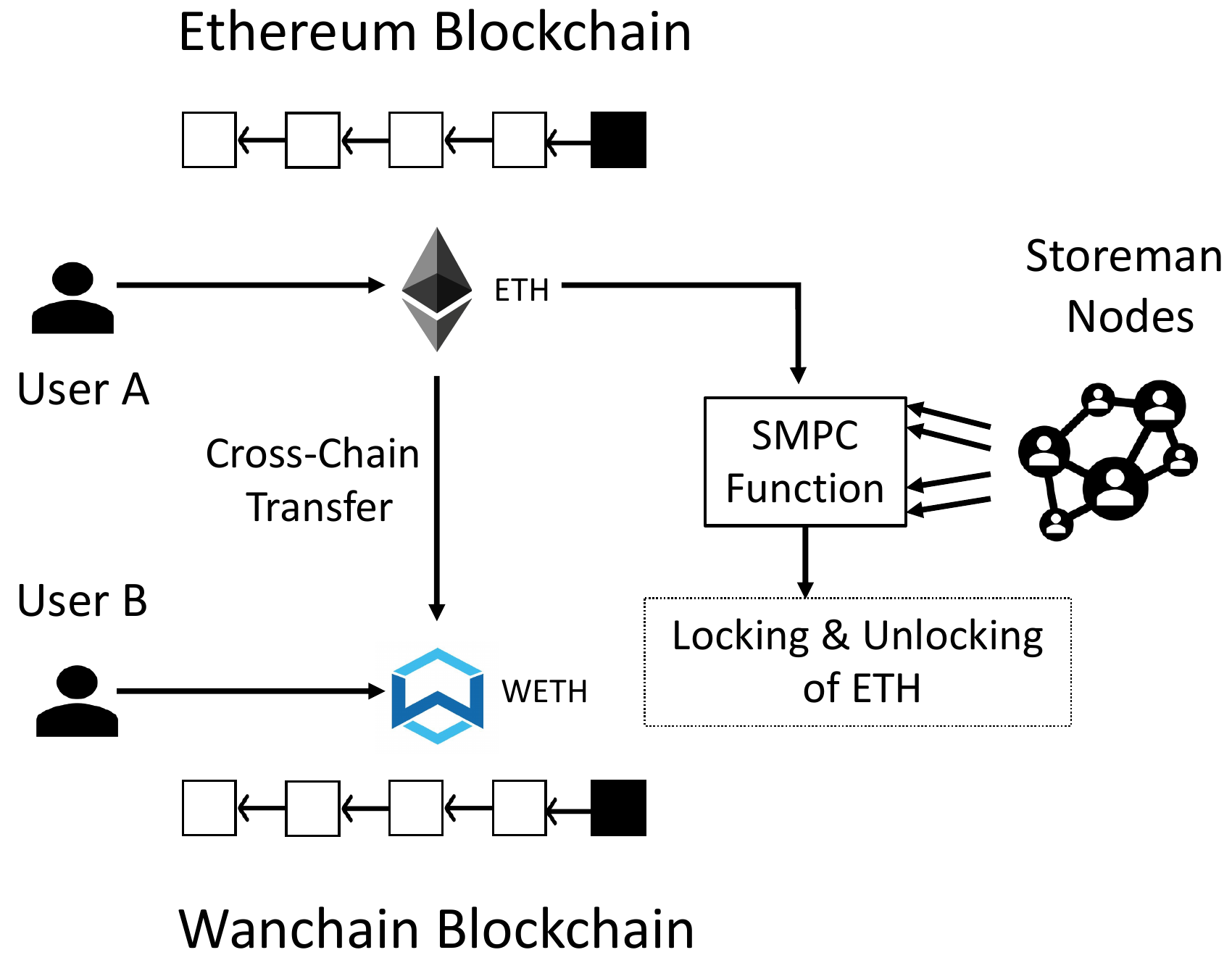}
        \caption{Cross-Chain transfer mechanism of blockchain using SMPC}
        \label{SMPC}
    \end{figure}

    An application of SMPC can also be seen in the Wanchain~\cite{Wanchain} Cross-Chain network. Figure~\ref{SMPC} reflects the SMPC idea in cross-chain transfer model. In Wanchain network, if user A wants to send an asset (say ETH) from one blockchain (say Ethereum blockchain) to user B on Wanchain blockchain,  then at first the asset value is locked in an account on its original chain using smart contract. This locked account holds control of the funds. The equivalent token WETH is sent to another user B of the Wanchain network. When user B wants to convert its WETH to ETH, the locked amount is released from the locked account and sent to user B, and the equivalent portion of WETH is burned. These locking and unlocking of asset value (ETH) happen using SMPC. Wanchain has a concept of Storeman nodes which work together and perform locking and unlocking of account. These Storeman nodes jointly work together to create public and private key pair of the related locked account. This shared account private key is scattered among the Storeman nodes as pieces of the key. To unlock the account, M out of N (M $\leq$ N) Storeman nodes contribute their shares of the private key to generate the signature using MPC jointly. 
    
     \subsection{Secret Sharing} In this concept, a secret is divided into multiple parts among the participants, and it is reconstructed by using a minimum number of parts. These parts are called shares and they are unique for each participant. Secret sharing is used to secure sensitive information. Secret sharing scheme is advantageous in SMPC for distributing the shares among parties. Shamir's secret sharing~\cite{Shamir} is already being used to distribute transaction data, without a significant loss in data integrity in blockchain~\cite{SSB}. Decentralized Autonomous Organizations (DAO) can take advantage of secret sharing by distributing the shares of information among the system nodes rather than storing full information in each node. Secret sharing in DAO can be practiced in consensus where each participating node stores a set of shares of the system state rather than storing full system state. These shares are points on polynomials which make up part of the state. 
     
      \begin{figure}[htbp]
        \centering
        \includegraphics[width = 0.7\textwidth]{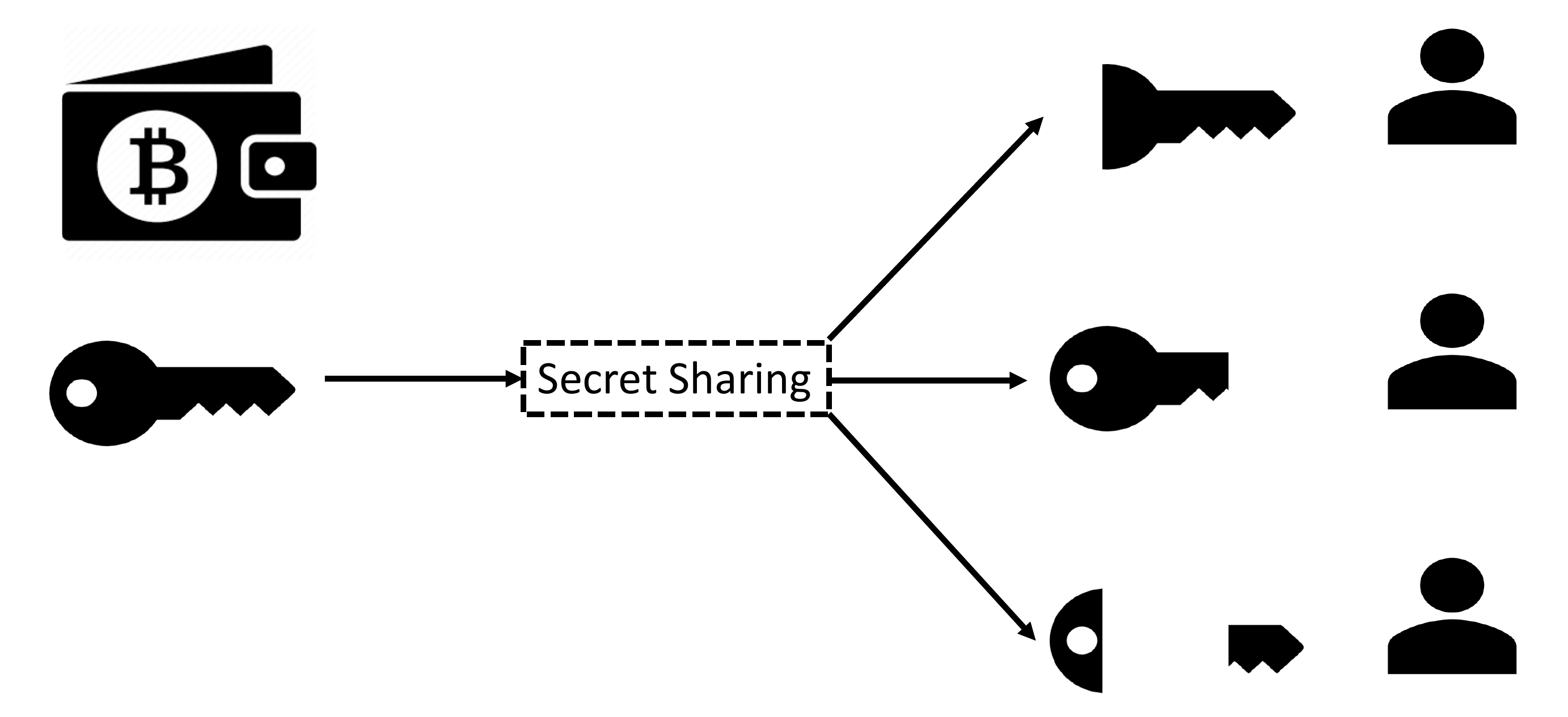}
        \caption{Secret-Sharing-Scheme 2-of-3 for a cryptocurrency wallet private key}
        \label{Secret-Sharing}
    \end{figure}
     
     Secret sharing schemes are also used in different off-chain and on-chain bitcoin wallets to safeguard the private keys of the crypto holders.
     For example, suppose an organization wants to store its bitcoin with a single master private key. In that case, secret sharing scheme helps to store the same key among multiple people. A simple example of this scenario will be sharing a bitcoin wallet key among three people by distributing the shares of the key. These individual shares do not convey any information about the actual key. However, any 2 of 3 people can reconstruct the key using their shares as presented in Figure~\ref{Secret-Sharing}. Secret sharing schemes can benefit blockchain by storing secret information in a decentralized way so that unauthorized parties cannot access it. Secret sharing is used in blockchain for different purposes such as secret share-based fair and secure voting protocol (SHARVOT)~\cite{SHARVOT} and new cryptocurrency based on mini blockchain~\cite{MiniBlockchain}.

\subsection{Commitment Scheme} A commitment scheme is a digital analog of a sealed envelop. It is a two-phase game between two parties where the phases are commit and open. Commit phase involves hiding and binding of a secret by the first party and send it to the second party; while open is to prove that the first party did not cheat the second party in the commit phase. Therefore, a commitment scheme satisfies the aforementioned two security properties: \textit{hiding} and \textit{binding}. Hiding ensures that the receiver cannot see the message before the open phase, while binding ensures that the sender cannot change the message after the commit phase. The following example shows a binding commitment: 
  \begin{enumerate}
      \item Pick a secret value $s$ to commit from $0$ to $p-1$ where $p$ is a large prime number;
      \item Calculate the value $c = g^s \mod \ p$;
      \item Publish the value $c$ as a commitment.
  \end{enumerate}
In the above example, the binding property follows as it is infeasible for the sender to find any other value $y$ which gives the same $c$. Here finding the value $s$ from known $c, p$ and $g$ is a computationally hard problem of discrete logarithm but any party can verify the commitment value $c$ if $s$ is provided.  There are many commitment schemes such as Pedersen commitment~\cite{Pedersen} and elliptic curve Pedersen commitment. Zerocoin~\cite{Zerocoin} uses Pedersen commitment to bind a serial number $s$ to Zerocoin $z$. The commitment $c$ is given as follows:
          $$c = g^s h^z \mod p.$$
Here $ g,h$, and $p$ are known to everyone, and the user chooses $s,z$ and computes and publishes the commitment $c$. These $s,z$ cannot be computed from $c$ even if one is provided. As a consequence, in Zerocoin when the serial number $s$ is published, the user can prove his/her ownership by providing $z$. Pedersen commitment has also been used to build blockchain-oriented range proof system, Bulletproof~\cite{Bunz} and its elliptic curve version is also successfully implemented in Monero~\cite{Maxwell}. A switch commitment scheme is designed for confidential transactions in blockchain~\cite{Switch}.
     
    \subsection{Accumulator} 
\begin{figure} [h]
        
        \centering
        \includegraphics[width = 0.60\textwidth]{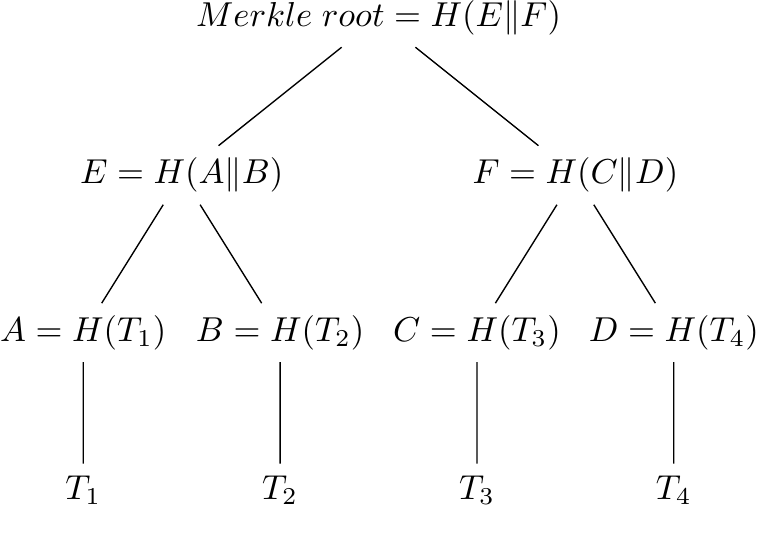}
        \caption{Merkle tree of blockchain transactions.}
        \label{Merkle tree}
    \end{figure}

An accumulator is a one-way function which gives a membership proof without revealing individual identity in the underlying set. This can be used in blockchain to build other cryptographic primitives such as commitment, ring signatures, and zero-knowledge proofs. Merkle tree, used in many cryptocurrencies, fits under a more comprehensive class of cryptographic accumulators which is space and time efficient data structure to test for set membership. Figure~\ref{Merkle tree} shows how blockchain transactions are represented in the Merkle tree, and the Merkle root is stored in the block structure of the blockchain. Non-Merkle accumulators are classified as RSA accumulators and elliptic curve accumulators.

In Zerocoin~\cite{Zerocoin}, an accumulator $A$ is computed by the network overall coin commitments $(c_1,c_2,\ldots,c_n)$ along with the appropriate membership witnesses for each item in the set. The witness $w$ is computed by the accumulation coins with the exception of one. In this way, during Zerocoin spend transaction, a user proves the knowledge of one coin by using that witness. This witness $w$ and accumulator $A$ are publicly verifiable without any trusted third party. Accumulator $A$ in Zerocoin is defined as:
        $$A = u^{c_1 \ c_2 \ c_3 \ \ldots \ c \ \ldots \ c_n} \mod \ N,$$
where the integers $A, u$ and $N$ are known to everyone.The coin $c$ is a Pedersen commitment of a coin serial number $s$ and the random number $z$. Zerocoin uses Random Number Generator (RNG) to generate different $s$ and $z$ to find $c$ using Pedersen commitment until $c$ is prime. The witness $w$ of a coin $c$ is defined as the accumulation of all coins with the exception of $c$:$$A = u^{c_1 \ c_2 \ c_3 \ \ldots \ c_n} \mod \ N.$$

Accumulators can also be employed for range proofs in blockchain. Accumulators are used in ~\cite{Accumulator} to design a stateless blockchain where in order to participate in consensus, the node only needs a constant amount of storage.
   
    \subsection{Oblivious Transfer (OT)} Oblivious Transfer is a two-party protocol between a sender $S$ and a receiver $R$. The general type of oblivious transfer is $k$-out-of-$n$ oblivious transfer $\binom{n}{k}$-$OT$, where $k < n$, in which $S$ holds $n$ messages and $R$ retrieves simultaneously $k$ of them without letting $S$ know about which $k$ out of $n$ messages $R$ received. Oblivious transfer is introduced by Rabin~\cite{rabin2005exchange} in which a sender sends a message to a receiver with probability $\frac{1}{2}$. The protocol is called as $\frac{1}{2}$-$OT$, and it is as following:
    \begin{enumerate}
        \item Sender S chooses two large primes $p,q$ and computes $N=pq$ and then the sender generates RSA public key $(e,N)$ such that $e$ is relatively prime to $(p-1)(q-1)$.
        \item S computes cipher text $c$ over message $M$ as $c = E_{(e,N)}(M) = M^e \mod N$ and sends $e, N, c$ to receiver R.
        \item R chooses a random $x \in \mathbb{Z}_{N^*}$ and sends $a = x^2 \mod N$ to $S$.
        \item S computes four square roots of  $a \mod N$ and chooses one of the roots $y$ at random and sends it to R.
        \item R checks whether $y^2\equiv a \mod N$ and if $y \not\equiv \pm x \mod N$, then R will be able to factor $N$ and, hence, be able to decrypt $c$ to recover $M$.
    \end{enumerate}
    
    $\frac{1}{2}$-$OT$ is \textit{complete} for secure multi-party computation. Oblivious transfer has been realized in secure multiparty computation to create private and verifiable smart contracts on blockchain~\cite{Raziel}. Oblivious transfer can also be utilized for exchange of secrets, private information retrieval, and building protocols for signing contracts. There has been loads of work done in oblivious transfer, and some of these works have been applied in blockchains such as Searchain~\cite{Searchain} and APDB~\cite{APDB} (for automated penalization of data breaches using crypto-augmented smart contracts). 

    \subsection{Oblivious RAM (ORAM)}  Oblivious RAM is a cryptographic protocol through which a client can safely store his/her data in an untrusted server. The client performs read and write operations remotely. ORAM hides the memory access pattern from the server as well as from outside entities accessing to that part of the data. Therefore, if a client performs two operations of equal length, then the polynomial-bounded adversarial server cannot distinguish between these operations. ORAM bestows freshness, confidentiality of data and integrity so it can be used in various blockchain use-cases and applications. Solidus~\cite{Solidus}, a protocol for confidential transactions on public blockchain, uses oblivious RAM. Solidus framework operates on a modest number of banks where each bank maintains a large number of user accounts.
    Solidus introduces a new primitive called Publicly Verifiable Oblivious RAM Machine (PVORM). Most of the previous usage of Oblivious RAM is performed by a single client to outsource storage. In Solidus, ORAM is used to store user account balances and uses PVORM to verify the valid transaction set of a bank. Oblivious RAM is also used in the client-server ORAM protocol~\cite{EVORAM}, Externally Verifiable Oblivious RAM, where Ethereum's automated crypto-currency contracts adjudicate the disputes occurred due to the malicious server by penalizing the server.
    
    \subsection{Proof of Retrievability (POR)}    With the advent of cloud computing, a client might outsource his/her data to the cloud, but still, the client needs a guarantee that the old data has not been modified or deleted. This can be achieved by using proof of retrievability~\cite{POR} which is an interactive mechanism between a client (verifier) and a server (prover) where the server provides a compact proof to the client that his/her data is intact and he/she can recover the data at any point of time. In this direction, to verify the proof, the client should be equipped with devices having some computational power and network access. This requirement hinders the large-scale adoption of POR by cloud users. To solve this issue, outsource proof of retrievability 
    (OPOR)~\cite{OPOR} is introduced where external auditors verify the POR with the cloud provider on behalf of the clients. OPOR protocol specification uses Bitcoin functionalities for the building blocks. 
    
    \textit{Permacoin}~\cite{Permacoin} uses proof of retrievability. The primary goal of Permacoin is the distributed storage of archival data. As in Bitcoin's mining mechanism, the client continuously invests his/her computational power, and in addition to the computational power, his/her storage is invested. As a consequence, Permacoin requires storage overhead and high bandwidth consumption. To solve these issues, \textit{Retricoin}~\cite{Retricoin} is proposed to repurpose the mining work in order to ensure the retrievability of a large file at any point of time. Retricoin also proposes a new algorithm for miners to mine collectively. Storj~\cite{Storj} also uses POR  to prove the existence of a fresh copy of a shard on the storer side. As a result, POR can be employed in many cryptocurrencies and blockchain applications.

\subsection{Post-Quantum Cryptography} 
Recent advances in quantum computing pose a severe threat to classical cryptography, as most of the widely used cryptography is based on the hardness of some problem which can be efficiently solved using quantum computers. Thus, research in the Post-Quantum cryptography~\cite{PQBC} has taken a massive leap. The security impact of breaking public key cryptography by quantum computers would be tremendous. Elliptic curve cryptography (ECC), which is an approach to public key cryptography, is mostly used in blockchain applications. Using a variant of Shor's algorithm~\cite{Shor}, a quantum computer can easily forge an elliptic curve signature that underpins the security of each transaction in blockchain and so breaking of ECC will affect blockchain in terms of broken keys, hence, digital signatures.  
    
Research in this field is in the rise to create Post-Quantum resistant digital signatures (BPQS)~\cite{BPQS} which is a hash-based signature and uses one-time signature (OTS) schemes as a building block. OTS does not depend on any number-theoretic hard problem, and it requires only a secure cryptographic hash function, hence, it is not vulnerable to Shor's algorithm. BPQS has advantages like shorter signatures, faster key generations, and customizable property. Post-Quantum cryptography is also used to design Post-Quantum blockchain~\cite{PQB} using one-time signature chains or to create secure crypto-currency based on Post-Quantum blockchain~\cite{PQBCurrency}. 

For the quantum proof solutions, research is now focused on Lattice-based cryptography ~\cite{LatticeBC}, multivariate cryptography~\cite{MultivariateBC}, hash-based cryptography~\cite{PQBC}, and code-based cryptography~\cite{CodeBc}. Most of the developed primitives within these areas offer either signatures or public keys that are orders of magnitude bigger than the currently used ones, and that is really a hard research challenge that we formulate as:
\vspace{1.0mm}
\begin{researchproblem}\label{ResearchProblemPostQuantumCrypto}
Construct a new blockchain mechanism that has comparably efficient public key addresses and comparably small digital signatures as the currently used ones, but that is based on Post-Quantum cryptographic schemes.
\end{researchproblem}

\subsection{Lightweight Cryptography} Conventional cryptographic methods such as RSA and SHA256, work well on systems having reasonable memory and processing power, but these methods are not suitable for devices constrained with memory, physical size, and battery. Conventional cryptographic methods are challenging to implement in resource-constrained devices due to implementation size, large key size, throughput, speed, and energy consumption. Nevertheless, to solve these issues, lightweight cryptography has evolved. Lightweight cryptography targets sensor networks, embedded systems and other variety of resource-constrained devices such as IoT end nodes and RFID tags. Lightweight cryptography is simpler and faster than conventional cryptography but less secure (suffers from many attacks).  
    In IoT, embedded devices having sensors are interconnected through a public or private network. As these are resource-constrained devices, lightweight cryptography solves the issues of communication, memory, and power consumption, but still lacks security. To provide better security, blockchain can be used in conjunction with the sensor network.
    
    Reference~\cite{IoT_Security} reinforces our point to use lightweight cryptography and blockchain for IoT devices to improve security (confidentiality and integrity of IoT device data). A lightweight scalable blockchain (LSB)~\cite{LSB} is also introduced to improve IoT security and privacy. LSB uses a lightweight hash function and lightweight consensus algorithm in order to achieve scalability, security, and privacy. Blockchain is also used to cater security in electric vehicles, cloud and edge computing~\cite{EVCE} which use lightweight cryptographic primitives like lightweight symmetric key encryption.
    
    \subsection{Verifiable Random Function (VRF)} This cryptographic primitive~\cite{VRF} is a pseudorandom function which gives a public verifiable proof of its output based on public input and private key. In short, it maps inputs to verifiable pseudorandom outputs. VRFs can be used to provide deterministic precommitments which can be revealed later using proofs. VRFs are resistant to pre-image attacks unlike traditional digital signature. VRF is a triple of the following algorithms:
    \begin{itemize}
        \item \textit{KeyGen(r)$\rightarrow$(VK,SK)}. Key generation algorithm generates verification key \textit{VK} and secret key \textit{SK} on random input \textit{r}.
        \item \textit{Eval(SK,M)$\rightarrow$(O,$\pi$)}. Evaluation algorithm takes secret key \textit{SK} and message \textit{M} as input and produces pseudorandom output string O and proof $\pi$.
        \item \textit{Verify(VK,M,O,$\pi$)$\rightarrow$0/1}. Verification algorithm takes input as verification key \textit{VK}, message \textit{M}, output string \textit{O}, and proof $\pi$. It outputs 1 if and only if it verifies that \textit{O} is the output produced by the evaluation algorithm on input secret key \textit{SK} and message \textit{M}, otherwise it outputs 0.
    \end{itemize}

    In context of blockchain, many Proof of Stake blockchains use VRF to perform secret cryptographic sortition such that electing leader and committee as part of underlying consensus protocol. Proof of Stake blockchain protocols given in~\cite{VRF_POS} use VRF to elect block proposers and voting committee members. Algorand~\cite{Algorand} and Witnet network protocol~\cite{Witnet} also employ VRF to conduct secret cryptographic sortition. Ouroboros Praos~\cite{Ouroboros_Praos} uses VRF on current timestamp and nonce to determine whether a participant is eligible to issue a block. Dfinity~\cite{Dfinity} network is a decentralized cloud computing resource which uses VRF to generate stream of outputs over time. Thus, the usage of verifiable random function brings many advantages to be exploited in blockchain and opportunities for more research.

 \subsection{Obfuscation} Obfuscation is a way of transforming a program $P$ into a "Black-box" equivalent of the program $Q= O(P)$ so that $P$ and $Q$ give the same output when the given inputs are same. It should be hard to find out the internal logic or structure of the program once it is obfuscated. Obfuscation aims to make reverse engineering difficult by making the program unintelligible while preserving its functionality. Finding a perfect black-box obfuscation is mathematically impossible. Along these lines, a weaker solution is to find an "Indistinguishability Obfuscation" so that one cannot determine whether the generated output is from the original program or the obfuscated program. A very simplified example for understanding the Indistinguishability Obfuscation, is the following: There are two equivalent programs $P = x*(y+z)$ and $P' = x*y + x*z$. They are obfuscated such that we have $O(P)$ and $O'(P')$. We say that the obfuscated programs $O$ and $O'$ are indistinguishable if on a received output $o$ one cannot determine which of the programs $O,O'$ gave that output. 
 
\begin{figure}[ht]
        \centering
        \includegraphics[width = 0.75\textwidth]{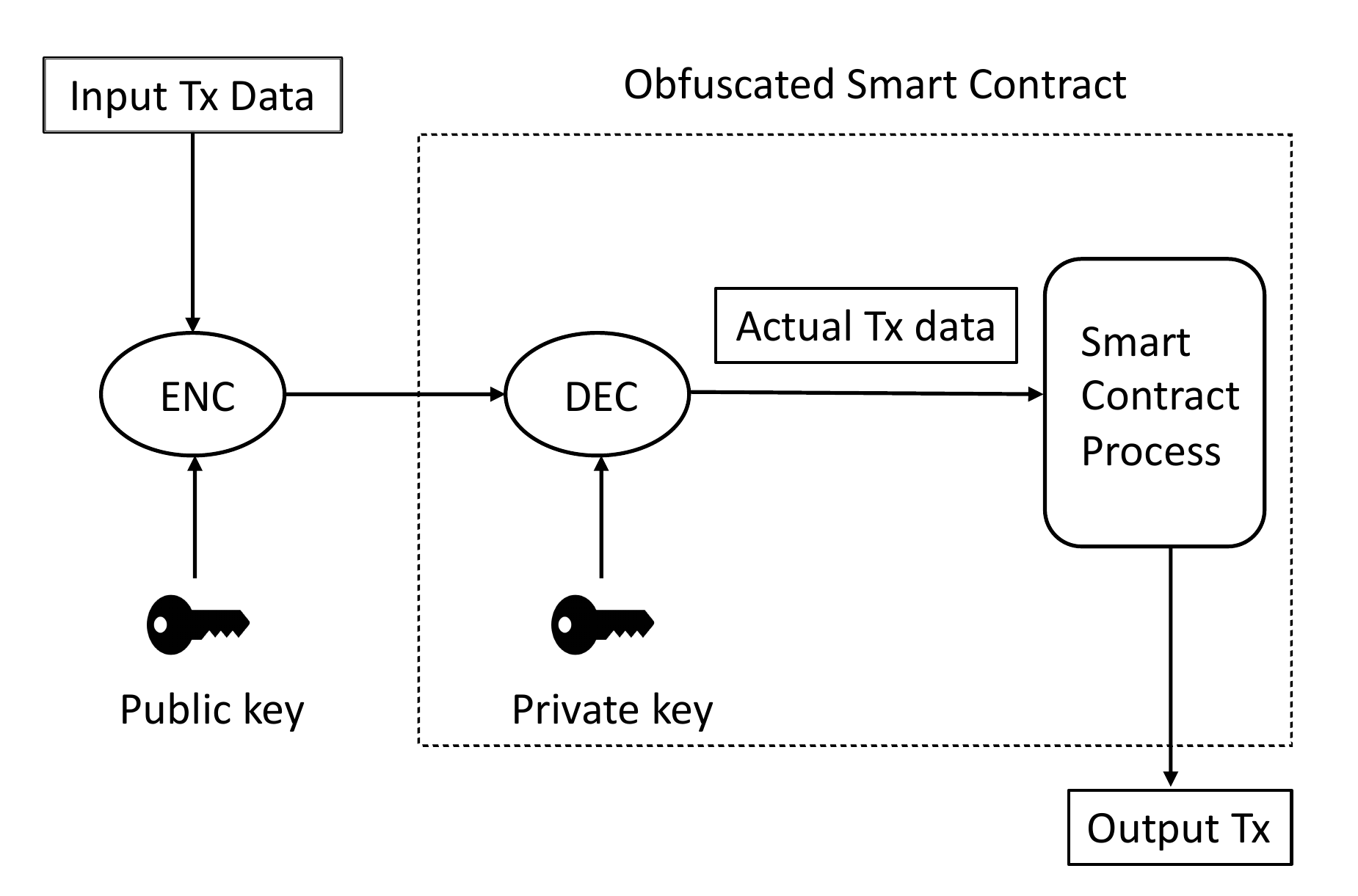}
        \caption{An example of smart contract obfuscation.}
        \label{Obfuscation}
\end{figure}

 Obfuscation can be applied for witness encryption, functional encryption, and restricted use of software. It can be applied in blockchain to turn smart contract into a black-box. An obfuscated smart contract can also possess a secret key to decrypt an encrypted input to the smart contract. As a result, publicly running contracts can possess secret data inside it by obfuscating the smart contract. Figure~\ref{Obfuscation} depicts an obfuscated smart contract which stores the private key corresponding to a public key which is used to encrypt the transaction data. It is hard to get the corresponding private key because of the obfuscated smart contract. 

One of the very first successful attempts to offer a very limited variant of obfuscation in Bitcoin was the standardization of the "Pay to script hash (P2SH) transactions"~\cite{P2SH}. The amounts of Bitcoins in P2SH transactions are sent to a script hash instead of a public key hash. We say that it was a limited variant of obfuscation because in order to spend Bitcoins received via P2SH, the recipient must provide a script that matches the script hash. Still, the successful acceptance of the P2SH transactions without causing a hard fork in Bitcoin showed that there is an interest in obfuscation in Blockchain, and that subject is a viable research area.
 
Research on obfuscation in Bitcoin~\cite{Obfuscation} has been conducted and can be compiled for other cryptocurrencies and blockchain applications. Obfuscation is also used in blockchain for power grid consumption~\cite{Obfuscation-Powergrid} where noise is added to the user's electricity consumption data through obfuscation, and the electricity consumption data is divided into random and non-random obfuscated data.

As noted in~\cite{Binstock2003} the definition and characteristics of some languages determine how easy is to obfuscate programs written in those languages. For example C, C++, Java and Perl are languages that offer easier program obfuscation. What about scripting languages used in Blockchain? We reformulate this question as a research problem:
\vspace{1.0mm}
\begin{researchproblem}\label{ResearchProblemObfuscation}
Study the easiness/hardness of obfuscating programs written in the scripting languages used in the current blockchain systems. Study the feasibility of applying some of the developed obfuscation techniques in C, C++, Java and Perl for the blockchain scripting languages.
\end{researchproblem}

    \section{Promising but yet not employed cryptographic primitives in blockchain}\label{NotUsedCryptoConcept} 
    This Section construes some cryptographic concepts which are promising candidates to be utilized in blockchain. These cryptographic concepts are not yet well-studied and fully applied in blockchain but constitute of some excellent properties which overlap with some desired properties of blockchain. Therefore, some use cases and blockchain services can benefit from these concepts. The included concepts in this Section have either not at all been studied for use in blockchain or have been studied but not implemented yet. We include references which show some initial ideas how to use  these concepts in blockchain, but these references do not give any details about concrete implementation. 

\subsection{Aggregate Signature} An aggregate signature allows creating a single compact signature from $k$ signatures on $k$ distinct messages from $k$ distinct signers. It provides faster verification as well as reduction in storage and bandwidth. As in blockchain, the requirement of storage and computation is high; aggregate signatures can be used for reduction in storage and computation. Aggregate signatures are the non-trivial generalization of multi-signatures (where all users sign the same message). There are two primary mechanisms of signature aggregation: general and sequential aggregation. In order to describe these mechanisms, assume a set of $k$ users having public-private key pair $(PK_i, SK_i)$ and user $i$ wants to sign message $M_i$.
\begin{enumerate}
        \item In general signature aggregation scheme, each user $i$ (from the group of $k$ users) creates signature $\sigma_i$ on his/her message $M_i$. Now to create aggregate signature, anyone can run public aggregation algorithm to take all $k$ signatures $\sigma_1, \sigma_2, \ldots ,\sigma_k$ and compress them into a single signature $\sigma$.
        \item In sequential signature aggregation scheme, user 1 signs $M_1$ to obtain $\sigma_1$; user 2 then combines $\sigma_1$ and $M_2$ to obtain $\sigma_2$; and so on. The final signature $\sigma$ is generated by user $k$ which binds $M_k$ and the signature $\sigma_{k-1}$. Sequential signature aggregation can only take place during the signing process.
\end{enumerate}
    
Techniques for aggregating signatures are known for a variety of signature schemes such as DSA, Schnorr, pairing-based, and lattice-based. Aggregate signature schemes should restrict any adversary from creating a valid aggregate signature on his/her own. Aggregate signatures have been proposed for Bitcoin~\cite{AggSign}, and they can be applied to other cryptocurrencies and blockchain designs. 
\vspace{1.0mm}
\begin{researchproblem}\label{ResearchProblemAgregateSignatures}
Construct an efficient new signature scheme based on aggregate signatures, that is specifically tailored for blockchain transactions.
\end{researchproblem}

\subsection{Identity-Based Encryption (IBE)} \label{IBE} Identity-Based Encryption first proposed as idea in \cite{ShamirIBE} and later realized as complete cryptographic primitive in \cite{IBE}, allows the encrypting party to use any known (or supposedly known) identity of any receiving party as its public key. Upon receiving the encrypted message, the receiving party asks a trusted third party "Private Key Generator (PKG)" to generate the corresponding private key. Then the receiver decrypts the message using the private key received by PKG. Nowadays, by using identity-based encryption, public keys can be generated using the social identities (Facebook, Twitter, LinkedIn). 
    
There are many flavors and extensions of IBE such as Hierarchical IBE~\cite{boneh2005hierarchical}, Attribute-based encryption~\cite{goyal2006attribute}, Decentralized attribute-based encryption~\cite{lewko2011decentralizing}, Functional encryption~\cite{agrawal2013functional} to name a few. 
    
One of the specifics of IBE is that it replaced the role of the Public-Key Infrastructure with the trusted third party PKG. The presence of a trusted third party somehow defeats the purpose to use it in permissionless blockchain, but still there is a scope to use it in the distributed ledger. Namely, it seems that IBE can be used in permissioned blockchain network. In permissioned blockchain a consortium of trusted third parties that distribute the private keys to the users can take the role to be IBE PKG. Another variant could be a smart contract layer being responsible for the generation of public-private key pairs inside the PKG using IBE. 
    
We identified that the use of IBE within blockchain has started in \cite{BAVP} as well as in supply chain management~\cite{BLIC}.. Still, there are a lot of challenges and opportunities for other blockchain applications and services. 
\vspace{1.0mm}
\begin{researchproblem}\label{ResearchProblemIBE}
Construct an IBE based (or IBE related) permissioned blockchain network.
\end{researchproblem}

\subsection{Verifiable Delay Function (VDF)} Verifiable Delay Function (VDF) is a function $f : X \rightarrow Y $ which takes a prescribed number of sequential steps to compute; however, the output can be easily verifiable by anyone. This delay function prevents malicious miners from computing the random output, and it also provides a short proof which is used during the verification of the output along with previously generated public parameters. Boneh et al. described the concept of VDF~\cite{VDF} as well as illustrated the idea about how it can be applicable to blockchain. VDF can be efficiently used as a way to add a delay in decentralized applications. VDF can be used in the application of decentralized systems such as in leader election process of consensus mechanisms, constructing randomness beacons and proofs of replication.  
    
    Delay function was initially implemented in Ethereum prototype~\cite{POD} where the main idea was verification of delay functions through smart contract by using a multi-round protocol. After this prototype implementation, the concept of verifiable delay function was proposed by Boneh et al. Nowadays several blockchain industries are trying to use VDF in their consensus mechanisms. Chia Network~\cite{Chia} which is open source blockchain is trying to use VDF in its ``Proof of space and time" consensus mechanism. Ethereum is also trying to develop a pseudorandom number generator using VDF. In this way, VDF brings opportunities to dig deeper and to be applied in the blockchain domain.
        
 \vspace{1.0mm}
\begin{researchproblem}[\cite{boneh2018survey}]\label{ResearchProblemVDF}
Finding a post-quantum secure simple VDF for the use of blockchain.
\end{researchproblem}

\subsection{Private Information Retrieval (PIR)} It is a cryptographic primitive in which a client queries to a server and retrieves the corresponding response from the server without exposing query terms as well as response. It is a weaker version of $1$-out-of-$n$ oblivious transfer. It can facilitate private blockchain queries to fetch transaction data privately from blockchain. Accordingly, it can be used to find out whether a particular transaction has been appended in the blockchain or can be used to check the transactions associated with the set of public keys and find out the remaining balances. In addition, PIR can be helpful to query transaction data in simplified payment verification (SPV) clients without compromising privacy. PIR requires an adequate amount of processing, but in the future there might be efficient PIR techniques which can be implemented in blockchain. PIR has also been applied in distributed storage~\cite{PIRDS} which can be further investigated and adopted in blockchain.
    
Paper \cite{PIR} sets several research problems in the area of blockchain transactions privacy and private information retrieval. We rephrase some of the research challenges postulated there:
\begin{researchproblem}[\cite{PIR}]\label{ResearchProblemPIR1}
Develop protocols where non-anonymous users can publish transactions that cannot be linked to their network addresses or to their other transactions.
\end{researchproblem}
\begin{researchproblem}[\cite{PIR}]\label{ResearchProblemPIR2}
Develop protocols where non-anonymous users can fetch details of specific transactions without revealing which transactions they seek.
\end{researchproblem}
\begin{researchproblem}[\cite{PIR}]\label{ResearchProblemPIR3}
Develop efficient and scalable protocols for anonymous publishing on permissioned blockchains, by combining the asynchronous Byzantine-tolerant consensus protocols for agreeing on transactions with the process of mixing users' announcements.
\end{researchproblem}

\subsection{Decentralized Authorization} Authorization and/or hiding sensitive data and actions are essential concepts of resource sharing in open and collaborative environments such as the Internet. Furthermore, in a decentralized form of authorization, parties have full control over their resources and authority to delegate it whether entirely or in part to other parties. An authorization system should provide only as little access to the users as possible to perform their jobs. 

Traditional access control is a centralized authorization server which imposes a problem of single point of failure. The centralized authorization scheme has different methods of authorization such as access control list or role-based access. 

In comparison, decentralized authorization is more efficient and easier in terms of time, resource and quality. A decentralized authorization system should be well administrated to give access privileges to the users. On the negative side, having in mind that the auditing is also a key component of authorization, in a decentralized manner, it is hard to efficiently implement it and to enforce it.
     
By using blockchain smart contract, some decentralized authorization systems have been designed, e.g., BlendCAC~\cite{BlendCAC} and WAVE~\cite{WAVE}. WAVE introduces an authorization layer for the name spaces and resources. Moreover, for the outside entities, a delegation of trust is used to obtain permission on a resource. Decentralized authorization and blockchain can be used to grow each other by combining one another in a specific way.
\vspace{1.0mm}
\begin{researchproblem}\label{ResearchProblemDecentralizedAuthorization}
Construct a decentralized authorization protocol for permissioned\\
blockchain that will provide access privileges as well as a delegation of these access to the users.
\end{researchproblem}

\subsection{White-Box Cryptography} White-box attack is a threat model where the attacker has full visibility of the internal data flow and can modify the data and code. White-box cryptography~\cite{chow2002white} aims to address the challenge of implementing a cryptographic algorithm in software in such a way that cryptographic assets remain secure even when subject to white-box attacks. A white-box cryptographic implementation must resist black-box (the attacker has access to only input and output of algorithm), grey-box (side-channel), and also white-box attacks. White-box cryptography is a way to implement cryptographic algorithms like RSA and AES so that the keys remain hidden all the time even during the execution. In some white-box implementations, the key is baked into the code and further concealed to use it in a cryptographic algorithm. In blockchain, it can be used to hide the private key inside the smart contract, and that key can be unlocked when smart contract executes and further it can be used to create a signature. 
     
     White-box cryptography can be orchestrated in blockchain to establish trust and privacy of assets. As in blockchain, key and seed secrets are a single point of compromise; these are the highly vulnerable and lucrative targets when stored in memory. To safely store the key, it can be obfuscated in white-box cryptography and further used for encryption/decryption. The implementation of white-box cryptography should be strong enough to facilitate the key storage in blockchain. It has been used in runtime self-protection in a trusted blockchain-inspired ledger~\cite{White-box} and can be promoted in other blockchain applications and services.

\subsection{Incremental Cryptography} The idea behind incremental cryptography~\cite{IncCrypto} is if there is a modification to some document $M$ to $M'$, then the time to update the result upon modification of $M$ should be "proportional" to the "amount of modification" done to $M$. Incremental cryptography can be used in incremental collision free-hashing or incremental digital signature. The initial idea proposed for incremental cryptography uses the example of a digital signature. The idea was to have a digital signature which is easy to update upon the modification of the underlying message. Suppose $M$ is a message and $\sigma$ is the corresponding signature. If $M$ is changed to $M'$ by adding/deleting any block, then the time to update the signature from $\sigma$ to $\sigma'$ should be "proportional" to the "amount of modification" done to get $M'$ from $M$. 

A proposal for construction of an incremental hash function based on SHA-3 is given in \cite{mihajloska2015reviving}, and a private blockchain "Kadena"~\cite{Kadena} proposes the use of either Merkle tree or incremental hashing for transaction verification. The concept of incremental hashing in Kadena blockchain is to update the distributed log among the blockchain nodes. 
\vspace{1.0mm}
\begin{researchproblem}\label{ResearchProblemIncrementalCrypto}
Construct a new blockchain mechanism that uses an incremental hash function for updates of the distributed ledger.
\end{researchproblem}

\subsection{Identity-based Broadcast Encryption (IBBE)} IBBE scheme~\cite{IBBE} can be considered as a generalization of identity-based encryption scheme (Section~\ref{IBE}) where instead of having one receiver, there are multiple receivers. In broadcast encryption the users are recognized by their identities rather than by their public keys. In a multi-receiver setting, IBBE proves as a powerful method to provide data security and privacy. In this scheme, a sender broadcasts the encrypted message to an intended set of users called privilege set. There can be many privilege sets with different cardinalities. A revocable IBEE scheme~\cite{RIBBE} shows a scenario of IBEE in which the involved players are the key authority, revoked and non-revoked users. In this setting, the decryption key is updated through the release of a key update material by the key authority. These decryption keys are updated only for the non-revoked users. In this scheme, a membership is revoked for a user if he/she is found malicious or his/her keys are compromised. This RIBBE scheme is further implemented in Charm framework~\cite{Charm}.
    
As blockchain is a multi-receiver setting, IBBE can be a propitious candidate to provide transaction data security and privacy. It can also be used in a permissioned blockchain to certify blocks of membership operation logs. RIBBE scheme as being very efficient in terms of computational complexity and communication can work efficiently as well in the case of blockchain.
\vspace{1.0mm}
\begin{researchproblem}\label{ResearchProblemIBBE}
Develop protocols to certify the blocks of membership operation logs in permissioned blockchain setting.
\end{researchproblem}

    \subsection{Other Techniques:}
    \begin{enumerate} 
        \item \textit{Message Authentication Code (MAC)}: It is a short piece of information (known as a tag) to authenticate a message which states that the message comes from the stated sender and it has not been changed. It can be used in blockchain to provide integrity of smart contracts or network data. A blockchain-based system for secure mutual authentication (BSeIn)~\cite{BSeIn} uses MAC for the authentication.
        
        \item \textit{Non-Interactive Witness Indistinguishability (NIWI)}: These are proof systems which are weaker variants of Non-Interactive zero-knowledge (NIZK) proofs. Witness Indistinguishable property states that the verifier cannot distinguish which witness is used to prove the statement by the prover, considering the case of existence of many witnesses. NIWI has been used to construct NIZK over POS based blockchain protocol~\cite{NIWI_NIZK} as well as recently, a new construction of publicly verifiable NIWI proofs from blockchain~\cite{NIWI} is also proposed. Hence NIWI proofs bring a new direction to be exploit within the blockchain domain.
        
        \item \textit{Position-based Cryptography}: In this cryptographic protocol~\cite{PBC}, the identity or the credentials of a party are derived from his/her geographical location. These credentials can be further used for position-based secure communication and position-based authentication. Position-based cryptography has not been applied in blockchain yet, but it looks promising.
        
        \item \textit{Elliptic Curve Diffie-Hellman Merkle (ECDHM) addresses}: These addresses~\cite{ECDHMAddress} can be used to exchange messages privately in the blockchain. It can be used in blockchain for secure communication among parties. ECDHM address is shared between the sender and the receiver as secret shares, and they use this shared secret to derive anonymous transacting addresses of each other. This address may only be exposed once they have the share to construct these addresses. In this way, it can be used for the privacy of transaction data.
        
        \item \textit{Verifiable Secret Shuffle}: It is a variant of a zero-knowledge proofs (an honest-verifier zeroknowledge) proposed in \cite{neff2001verifiable}. An initial application of verifiable shuffles has been proposed as a mixing service for Ethereum~\cite{cryptoeprint:2019:341}.
        
    \end{enumerate}

\section{Conclusion} \label{Conclusion}
The goal of this work was to offer a systematic study of available cryptographic concepts and to identify different research directions and problems. 
Based on these reviewed concepts and associated properties, we hope that the paper will help cryptographers interested in blockchain to choose a challenging research problem and for practitioners to choose a suitable concept for their particular use case.

Current transitions to blockchain enabled solutions by different  industries give rise to more research on this technology. Academic and industrial research is focused on making blockchain cost efficient in terms of computational power, memory requirements and security. Many existing cryptographic concepts have been embraced for blockchain use. This paper systematizes the current state-of-the-art knowledge of existing cryptographic concepts used in the blockchain. It also gives a brief description of the used cryptographic concept and points to the available blockchain models that are using that concept. The paper also identifies some concepts which have not yet been used in blockchain but can be beneficial if applied in the blockchain. Apart from existing cryptographic concepts, the paper also presents the basic building blocks of blockchain and how these building blocks are dependent on each other.

Table \ref{Techniques-table} summarizes all of the cryptographic concepts (used or with potentials to be used in blockchain) presented in this work.

\bibliographystyle{IEEEtran}
\bibliography{reference.bib}

\end{document}